\documentclass[12pt,singlespacing,a4paper,oneside]{book}

\usepackage{mathpazo}
\usepackage{longtable}
\usepackage{array}
\newcolumntype{L}[1]{>{\raggedright\let\newline\\\arraybackslash\hspace{0pt}}m{#1}}
\newcolumntype{C}[1]{>{\centering\let\newline\\\arraybackslash\hspace{0pt}}m{#1}}
\newcolumntype{R}[1]{>{\raggedleft\let\newline\\\arraybackslash\hspace{0pt}}m{#1}}

\usepackage[english]{ing-thesis/ingthesis}
\usepackage[latin1]{inputenc}
\usepackage{graphicx}
\usepackage{subfigure}
\usepackage{latexsym}
\usepackage{amsmath}
\usepackage{amsthm}
\usepackage{amssymb}
\usepackage{color}
\usepackage{xcolor}
\usepackage[hidelinks=true]{hyperref}
\usepackage{textcomp}
\usepackage{package/3Talg}
\usepackage{algorithm}
\usepackage[noend]{algpseudocode}
\usepackage{xspace}
\usepackage{hyperref}
\usepackage{comment}
\usepackage{multirow}
\usepackage{listings}

\usepackage{doc}

\usepackage[toc,page]{appendix}

\makeatletter

\g@addto@macro\UrlBreaks{\do\-}	% For URL references

\newcommand{\labitem}[2]{%
	\def\@itemlabel{\textbf{#1}}
	\item
	\def\@currentlabel{#1}\label{#2}}
\makeatother
\newcommand{\itcaption}[1]{\caption{\textit{#1}}}

\newcommand{\remove}[1]{}
\newcommand{\textbi}[1]{\textit{\textbf{#1}}}

\newcommand{\ite}{\begin{itemize}}
\newcommand{\eti}{\end{itemize}}
\newcommand{\enu}{\begin{enumerate}}
\newcommand{\une}{\end{enumerate}}

\newcommand{\figref}[1]{Figure~\ref{#1}}
\newcommand{\tabref}[1]{Table~\ref{#1}}

\renewcommand{\baselinestretch}{1.25} 

\makeatother
\newcommand{\property}[1]{\textbi{#1}}

\newif\ifshowcomments
\showcommentstrue

\ifshowcomments
\newcommand{\mynote}[2]{\fbox{\bfseries\sffamily\scriptsize{#1}}
	{\small$\blacktriangleright$\textsf{\emph{#2}}$\blacktriangleleft$}}
\else
\newcommand{\mynote}[2]{}
\fi

\definecolor{amber(sae/ece)}{rgb}{1.0, 0.49, 0.0}
\definecolor{silver}{rgb}{0.75, 0.75, 0.75}
\definecolor{eclipseBlue}{RGB}{42,0.0,255}
\definecolor{eclipseGreen}{RGB}{63,127,95}
\definecolor{eclipsePurple}{RGB}{127,0,85}
\definecolor{backcolour}{rgb}{0.95,0.95,0.92}
\definecolor{dollarbill}{rgb}{0.52, 0.73, 0.4}
\definecolor{alizarin}{rgb}{0.82, 0.1, 0.26}

\lstdefinelanguage{MyLang}
{
  morekeywords=[1]{
    import, bool, Option, String, pub, Vec, usize,
    if,
    while,
    for
  },
  morekeywords=[2]{false, Some, None, true, Mining},
  morekeywords=[3]{is_some, otry!},
  morekeywords=[4]{byzantine, self},
  sensitive=false, % keywords are not case-sensitive
  morecomment=[l]{//}, % l is for line comment
  morecomment=[s]{/*}{*/}, % s is for start and end delimiter
  morestring=[b]" % defines that strings are enclosed in double quotes
}

\newtheorem{theorem}{Theorem}
\newtheorem{corollary}{Corollary}

\begin{document}

\frontmatter

\setcounter{tocdepth}{2}

\titlepage{
\begin{figure}
\begin{center}
\includegraphics[width=8cm]{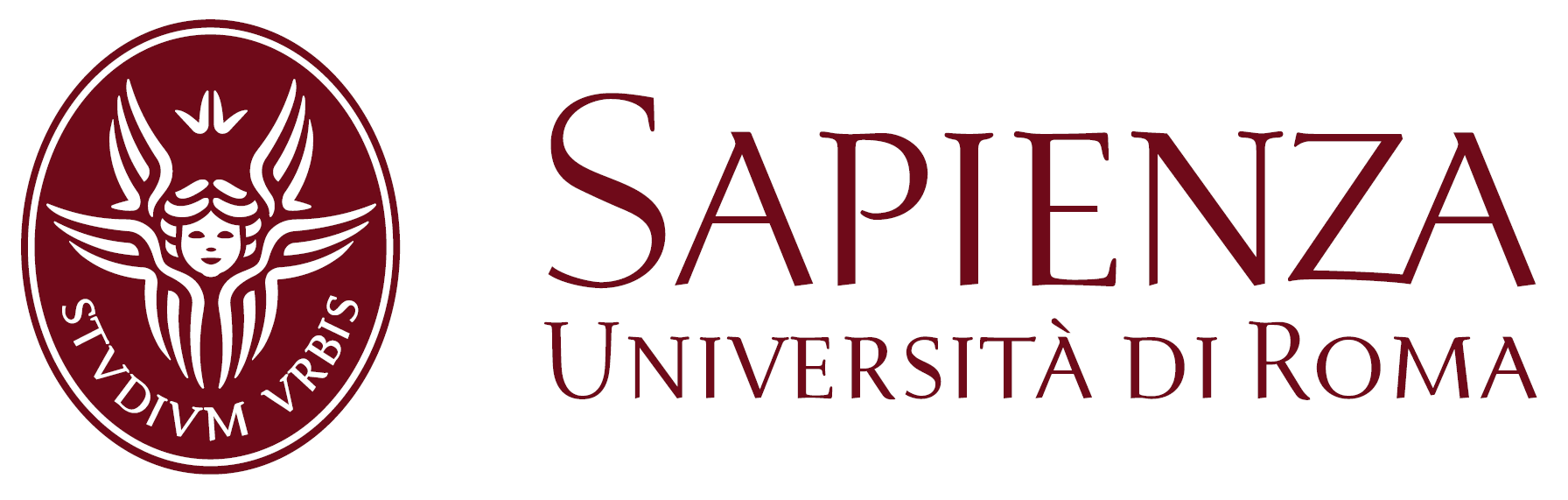}
\end{center}
\end{figure}

\begin{center}
\textsc{{Faculty of Information Engineering, Computer Science and Statistics\\Department of Computer, Control and Management Engineering}
\vspace{0.6cm} \Large{\\M.Sc. in Engineering of Computer Science\\}
\vspace{0.6cm} \normalsize{ }} \vspace{2cm}
\LARGE{Assessing Security and Performances of\\Consensus algorithms for Permissioned Blockchains}
\end{center}
\vspace{3cm}
\textsc{Author:} Stefano De Angelis \vspace{0.5cm}
\\ \\
\textsc{Advisor:} Roberto Baldoni \hspace{5.0cm}\textsc{Co-Advisor:} Leonardo Aniello \\
\textsc{Assistant Advisor:} Federico Lombardi \\}

\vspace{2.0cm}

\begin{center}
\textsc{Academic Year\\2016-2017}
\end{center}

%%%%%%%%%%%%%%%%%%%%%%
%% Author's Address %%
%%%%%%%%%%%%%%%%%%%%%%

\vspace*{\fill}
{
\noindent\underline{\hspace{16.48cm}}

\noindent{\textbf{Assessing Security and Performances of Consensus algorithms for Permissioned Blockchains.}\\M.Sc. Thesis\\ \smallskip © Stefano De Angelis. All rights reserved.\\}

\noindent Department of Computer, Control, and Management Engineering ``Antonio Ruberti''

\noindent Sapienza University of Rome

\noindent Via Ariosto 25, I-00185 Rome, Italy

\noindent \texttt{stefano.deangelis@me.com}

}

\cleardoublepage

%%%%%%%%%%
% Dedica %
%%%%%%%%%%
\newpage
\vspace*{6.5cm}
\begin{flushright}
{To my parents, who encouraged me\\to fly toward my dreams.\\}
\vspace{0.1cm}
{To my Family, who always supported me.}
\end{flushright}

%%%%%%%%%%%%%%%%%%
% RINGRAZIAMENTI %
%%%%%%%%%%%%%%%%%%
\cleardoublepage
\section*{Acknowledgements}
{For the fulfilment of this work and for the achievement of this important goal, I want to thank all the people who supported me in this period. First of all I would like to thank the professor Roberto Baldoni, for having interested me to the cyber security and for giving me the opportunity to carry out this thesis project abroad. Secondly, I would like to mention the professor Leonardo Aniello, for helping and supporting me during the entire work.

Then I would also like to thank the professor Vladimiro Sassone, for giving me a warm welcome to his research group at the University of Southampton, for his precious advices and for having believed in me. Moreover I would mention also the doctor Andrea Margheri, for his kindness and helpfulness in monitoring my job.

\smallskip
A special thank goes to Federico, for being a friend more than a colleague. For having encouraged me, also in difficult times. For helping me in the realisation of this work, as if it was his, addressing me step by step. For being with me during important decisions and for being always ready to give me a sincere and precious advise. Thank you.

Finally I would thank to all my family, because without them this goal would not be possible to achieve and, for their unconditional love. Then to all my friends to be always with me, even when I am far away from home.}

%%%%%%%%%%%%%
% Citazione %
%%%%%%%%%%%%%
\cleardoublepage
\vspace*{6.5cm}
\begin{flushright}
\emph{''The main advantage of blockchain technology is supposed to be that it's more secure, but new technologies are generally hard for people to trust, and this paradox can't really be avoided.''}

\smallskip
Vitalik Buterin
\end{flushright}

%%%%%%%%%%%%
% ABSTRACT %
%%%%%%%%%%%%
\cleardoublepage

\newpage
\rule{0pt}{55pt}
\begin{center}
\textsc{\textbf{Abstract}\\}
\end{center}
\begin{comment}

\begin{itemize}
\item Why blockchain is important: Interest in because its properties. Used first as distributed ledger for crtyptocurrency in trustless scenarios.
\item Why permissioned blockchains? Different consensus algorithms are rising, all of them different from the permissionless.
\item Due to the huge amount of new consensus algorithms, several problematics are rising all highlighted in the Cachin's paper.
\item Nell?ambito di questa problematica il nostro primo contributo riguarda l?analisi di due dei più importanti algoritmi per blockchain permissioned Aura and Clique, in termini di sicurezza e performance con l?aiuto del CAP Theoreme.
\item Esigenza di un benchmark comune, una sorta di framework comune per valutare .
\item Passaggio da modello formale ad implementazione pratica che non può coprire tu?tto in termini di sicurezza ma solo un sottoinsieme di aspetti da valutare tramite sperimentazione (Parte di cachin).
\item Secondo contributo, appunto, riguarda la definizione e il design di un benchmark per valutare diversi algoritmi di consenso per permissione blockchain sotto un modello generico sicurezza e performance di questi algoritmi.
\end{itemize}
\end{comment}

% Blockchain technology
\textit{Blockchain} is a novel technology that is rising a lot of interest in the industrial and research sectors because its properties of decentralisation, immutability and data integrity. 
Initially, the underlying consensus mechanism has been designed for \textit{permissionless blockchain} on trustless network model through the \textit{proof-of-work}, i.e. a mathematical challenge which requires high computational power. This solution suffers of poor performances, hence alternative consensus algorithms as the \textit{proof-of-stake} have been proposed. 

Conversely, for \textit{permissioned blockchain}, where participants are known and authenticated, variants of distributed consensus algorithms have been employed. 
However, most of them comes out without formal expression of security analysis and trust assumptions because the absence of an established knowledge. Therefore the lack of adequate analysis on these algorithms hinders any cautious evaluation of their effectiveness in a real-world setting where systems are deployed over trustless networks, i.e. Internet. 

\smallskip
In this thesis we analyse security and performances of permissioned blockchain. Thus we design a general model for such a scenario in a way to propose a general benchmark for the experimental evaluations. This work brings two main contributions.

The first contribution concern the analysis of \textit{Proof-of-Authority}, a Byzantine Fault-Tolerant consensus protocol. We compare two of the main algorithms, named \textit{Aura} and \textit{Clique}, with respect the well-established \textit{Practical Byzantine Fault-Tolerant}, in terms of security and performances. We refer the CAP theorem for the consistency, availability and partition tolerance guarantees and we describe a possible attack scenario in which one of the algorithms loses consistency. The analysis advocates that Proof-of-Authority for permissioned blockchains deployed over WANs experimenting Byzantine nodes, do not provide adequate consistency guarantees for scenarios where data integrity is essential. We claim that actually the Practical Byzantine Fault-Tolerant can fit better for permissioned blockchain, despite a limited loss in terms of performance.

The second contribution is the realisation of a benchmark for practical evaluations. We design a general model for permissioned blockchain under which benchmarking performances and security guarantees. However, because no experiment can verify all the possible security issues permitted by the model, we prototype an adversarial model which simulate three attacks, feasible for a blockchain system. We then integrate this attacker model in a real blockchain client to evaluate the resiliency of the system and how much the attacks impact performances and security guarantees. 
\cleardoublepage

%%%%%%%%%%
% INDICE %
%%%%%%%%%%
\tableofcontents

%%%%%%%%%%%%
% CHAPTERS %
%%%%%%%%%%%%
\mainmatter
\chapter*{Introduction}
\addcontentsline{toc}{chapter}{Introduction}
\label{c00:intro}
% !TEX root = ../../Tesi-1side.tex

%Into BC + application domain
Blockchain is one of the most disruptive technologies of the recent years. Firstly appeared as a decentralised public ledger for the Bitcoin cryptocurrency~\cite{bitcoin2008}, blockchain is nowadays widely exploited to a broad area of application domains. Its distinguishing properties of data immutability, integrity and full decentralisation are key drivers for general purpose exploitations, ranging from Cloud computing to business-to-business applications. 

%What is blockchain
Essentially, blockchain is a linked data structure replicated over a peer-to-peer network, where transactions are issued to form new blocks.  Peers achieve distributed consensus on transaction ordering by placing them into new blocks; each block is linked to the previous by means of its hash. Such block creation process is carried out by distinguished nodes of the network, named \textit{miners}, according to a distributed consensus algorithm. Besides cryptocurrencies \`a la Bitcoin, miners can also run \textit{smart contracts}, immutable programs deployed and executed in a decentralised fashion upon a blockchain. Ethereum \cite{wood2014ethereum} is the first smart contract framework of its denomination, whose transactions represent non repudiable executions of smart contracts. 

%Permissioned blockchain
Blockchain systems like Bitcoin and Ethereum are called \textit{permissionless}, i.e. any node on the Internet can join and become a miner. Distributed consensus is here achieved via so-called \emph{Proof of Work} (PoW), a computational intensive hashing-based mathematical challenge.  PoW enjoys strong integrity guarantees and tolerates a sheer number of attacks~\cite{Garay2015}, but this comes at a huge cost: lack of performance. Moreover the PoW-based blockchain achieve \text{eventual consensus} where conflicts between blocks are possible and eventual solved later by some architectural rules, i.e. eventual consistency. This has led, together with the absence of privacy and security controls on data, to so-called \textit{permissioned} blockchain, where an additional authentication and authorisation layer on miners is in place. Examples of permissioned blockchains are Multichain~\cite{multichain} and R3 Corda~\cite{corda}, (Hyperledger) Fabric~\cite{Cachin2016ArchitectureOT} and permissioned-oriented Ethereum clients. Such systems have prompted federation of companies thus to facilitate their interactions without giving out guarantees on control and computation of data. By way of example, the Cloud Federation-as-a-Service solution~\cite{faas} is exploiting a permissioned blockchain to underpin the governance of a federation of private Clouds~\cite{gaetani2017blockchain} connected via the Internet. 

%Consensus in Permissioned blockchain
Being the operating environment more trusted, permissioned blockchains rely on message-based consensus schema, rather than on hashing procedures. In such setting, dominant candidates are Byzantine Fault-Tolerant (BFT) algorithms such as the Practical BFT (PBFT)~\cite{castro1999practical}. Indeed, BFT-like algorithms have been widely investigated for permissioned blockchains~\cite{tseng2016recent} with the aim of outperforming PoW while ensuring adequate fault tolerance. Furthermore BFT-based blockchain offers, by design of the underling consensus schemas, strong consistency on transactions despite availability. However this shift to BFT consensus in blockchain comes with a set of challenges. Even if permissioned blockchain promise to offer better performances, there is a lot of confusion around this. 

% Thesis problem
Actually countless proposal have appeared, each one based on different ideas and assumptions. Most of these implementations do not have exhaustive security analysis and detailed documentations because there is not an accepted standard for consensus in blockchain \cite{cachin2017blockchains}. This poor informations generate ambiguity between current implementations and make difficult their evaluation and comparison.

\subsection*{Contributions}

The goal of this thesis is to clarify the context of permissioned blockchain. We focus on the underling consensus protocols and, by referencing the most general problems in distributed computing, we propose a solution to evaluate security and performance guarantees in a general blockchain framework.

The first contribution of the thesis is the analysis of two algorithms for blockchain permissioned. The algorithms, namely \textit{Aura} and \textit{Clique}, belongs to a new interesting family of BFT protocols, i.e. \textit{Proof-of-Authority} (PoA), and promise to offer better performances than the well known PBFT. Because the poor documentation, we understand the functioning of these algorithms through reversed engineering techniques. Therefore we analyse their security guarantees, by referencing the CAP Theorem, comparing the results with the well established PBFT protocol \cite{poaVSpbft}. This work has been presented at the conference ITASEC18.

The second contribution of the thesis is a benchmark proposal for permissioned blockchains. For this scope, we firstly define a standard model for such systems. Thus we design an attacker model to outline the possible vulnerabilities and attack scenarios. Then we propose the implementation guidelines for a Byzantine client, based on the attacker model, which simulate in real systems the established attacks. In such a way we propose to measure how the attacks impact to security and performance properties in such a scenarios.

\subsection*{Thesis structure}
In Chapter \ref{c01:context} we outline the context, by introducing the blockchain technology and the most famous implementations. In Chapter \ref{c02:consensus} we introduce the problem of consensus in distributed systems and in permissioned blockchains. In Chapter \ref{c10:poa} is presented the CAP Theorem evaluation of security and performances between two PoA algorithms and PBFT. In Chapter \ref{c04:benchmark} we introduce the benchmark evaluation, the system model for permissioned blockchain and the threat model. Finally in Chapter \ref{c05:conclusion} we summarise the obtained results and we outline the future works.

\chapter{Background and Context}\markchapter[Chapter~\thechapter:~Background and Context]
\label{c01:context}
% !TEX root = ../../Tesi-1side.tex
\section{Blockchain Technology} \label{s:blockchain}
\begin{figure}[pht]
	\begin{center}
		\includegraphics[width=11cm]{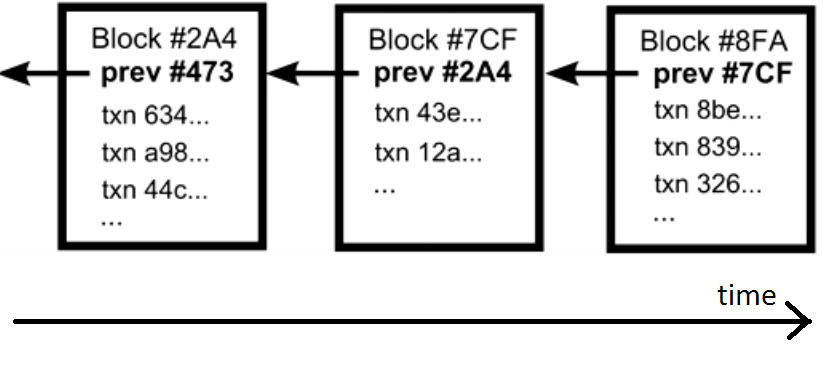}
	\end{center}
	\itcaption{Blockchain representation. Blocks reference each other by the hash of the previous}
	\label{fig:blockchain}
\end{figure}

%What is a Blockchain
Blockchain is a disruptive technology emerged in the recent years, firstly employed within Bitcoin's cryptocurrency~\cite{bitcoin2008} as public ledger. It is essentially a decentralised and distributed data structure, replicated over a peer-to-peer (p2p) network. It mainly consists of consecutive chained blocks, each one linked with the hash of the previous, containing records. These records witness transactions occurred among participants (see \figref{fig:blockchain}), such transactions represents an asset exchange between the nodes of the network, e.g. Bitcoin's cryptocurrency.

The process of block creation is called \textit{mining~process} and is carried out by special nodes called \textit{miners}, which periodically propose new blocks according to a distributed consensus algorithm. In such a way participants achieve transactions agreement, share the same state and guarantee consistency. Because decentralised and fully-replicated, blockchain also deals with the necessity of trust in a central authority as the most general modern systems do. There is not the necessity of a trusted third-party which controls all transactions and manage data, indeed every node can independently verify the consistency of the shared state. These important characteristic ensures data integrity and data immutability, since any improper alteration (e.g. tampering) would be immediately detected and rejected by the network. 

\smallskip
% Smart contract and Ethereum
Moreover, blockchains have the possibility to deploy and execute immutable programs, i.e. \textit{smart~contracts}, in a fully decentralised fashion. The first and most common framework for decentralised applications was proposed by Ethereum \cite{wood2014ethereum}.

%Permissionless and Permissioned blockchain
Blockchain systems like Bitcoin and Ethereum are so called \textit{permissionless}, i.e. any node on the Internet can join the network and become a miner. In this case distributed consensus is achieved via the so-called \emph{Proof of Work} (PoW), a computational intensive mathematical challenge. Conversely if there is an authentication and authorisation layer for miners, then the blockchain is \textit{permissioned}. 

\subsection{Public versus Private Blockchain} \label{s:blockchain-pub-priv}
Blockchain systems can be classified on the basis of the access types in different permission models. Participants of a blockchain network have rights to: (i) accessing data on the blockchain (\textit{Read}), (ii) submitting transactions (\textit{Write}) and (iii) updating the state with new blocks (\textit{Commit}) \cite{hileman2017bench}. 

The work proposed by BitFury and Garzik in \cite{publicVSprivate1} and \cite{publicVSprivate2}, define blockchain systems on the basis of these rights. Specifically a blockchain can be essentially divided in two main classes:
\ite
\item \textit{public blockchain}: no restrictions applied on Read operations;
\item \textit{private blockchain}: only a predefined list of entities is allowed to run Read operations.
\eti
Hence for the Write and Commit operations are identified two classes of blockchain:
\ite
\item \textit{permissionless blockchain}: there are no restrictions on Write and Commit operations;
\item \textit{permissioned blockchain}: only a predefined list of entities is allowed to Write and Commit operations.
\eti
In permissionless blockchain anyone can join the network, execute Write operations, Commit operations and, participate to consensus. Contrarily in permissioned blockchain nodes are known thanks to an authentication mechanisms, and only these nodes can control the blockchain.

\tabref{table:pub-priv} shows a comparison between four blockchain models based on these classes.

\begin{table}[!htb]
\centering
\begin{tabular}{c|c|c|c|}
\cline{2-4}
\multicolumn{1}{l|}{} & Read & Write & Commit \\ \hline
\multicolumn{1}{|c|}{\begin{tabular}[c]{@{}c@{}}Public\\ permissionless\end{tabular}} & Open to anyone & Anyone & Anyone \\ \hline
\multicolumn{1}{|c|}{\begin{tabular}[c]{@{}c@{}}Public\\ permissioned\end{tabular}} & Open to anyone & \begin{tabular}[c]{@{}c@{}}Authorised\\ participants\end{tabular} & \begin{tabular}[c]{@{}c@{}}All or subset of\\ authorised\\ participants\end{tabular} \\ \hline
\multicolumn{1}{|c|}{Consortium} & \begin{tabular}[c]{@{}c@{}}Restricted to an\\ authorised set of\\ participants\end{tabular} & \begin{tabular}[c]{@{}c@{}}Authorised\\ participants\end{tabular} & \begin{tabular}[c]{@{}c@{}}All or subset of\\ authorised\\ participants\end{tabular} \\ \hline
\multicolumn{1}{|c|}{\begin{tabular}[c]{@{}c@{}}Private\\ permissioned\end{tabular}} & \begin{tabular}[c]{@{}c@{}}Fully private or\\ restricted to a limited\\ set of authorised\\ nodes\end{tabular} & \begin{tabular}[c]{@{}c@{}}Network\\ operator only\end{tabular} & \begin{tabular}[c]{@{}c@{}}Network\\ operator only\end{tabular} \\ \hline
\end{tabular}
\caption{Main types of blockchain systems}
\label{table:pub-priv}
\end{table}

Public permissionless blockchains, e.g. Bitcoin, operate in a hostile environment, requiring the use of crypto-techniques to incentivise participants in behave honestly. Contrarily, private permissioned blockchains operate in an environment where participants are already known. In this scenario nodes are incentivised to behave honestly because are authenticated, so every misbehaviour can be detected and convicted. Consortium blockchains are a special class of permissioned blockchain where memebers of a consortium operate in a given context.

The private settings where participants are known and authenticated, make blockchain platforms step away from their original purpose and enter in the domain of database-replication protocols, notably, the classical \textit{state-machine replication} (SMR) \cite{schneider1990implementing}, where only a predetermined set of nodes participate to the network..

\section{Bitcoin: The Genesis of Blockchain} \label{s:bitcoin}
Bitcoin~\cite{bitcoin2008} was the first popularised application of permissionless blockchain for a virtual currency. It is an electronic payment system based on cryptographic proof, which avoid the need of a centralised authority, e.g. a bank. Current payment systems require trusted third-party to process transactions, however this mediation is often subject to transaction costs, limiting the minimum practical transaction size and cutting off the possibility for small casual transactions.

To guarantee machine-to-machine payments, the Bitcoin protocol uses a flood propagated p2p network that processes transactions in a fully decentralised system. Transaction order is guaranteed by an hash based consensus mechanism. To maintain a consistent, immutable, and therefore trustworthy, ledger of transactions, nodes share the Bitcoin Public Ledger \figref{fig:bit-ledg}, i.e. a blockchain with the list of all transactions ever made.

\begin{figure}[pht]
	\begin{center}
		\includegraphics[width=16cm]{figures/context/BitCoin-ledger}
	\end{center}
	\itcaption{Bitcoin Public Ledger.}
	\label{fig:bit-ledg}
\end{figure}

Transactions are managed by asymmetrical cryptograph as a chain of digital signatures. Each owner transfers the coin by digitally signing a hash of the previous transactions in input and the public key of the next owner and adding these to the end of the coin as output. In this way only users which have previously received coins, can process new transactions. In Bitcoin, every node has a balance of owned cryptocurrency coins identified by the history of previous transactions.  

When a node receive a transaction verifies the crypto-currency balance of the sender by checking his previous unused transactions with its available balance. In such a way only valid transfers are accepted. Moreover, Bitcoin protocol prevents also the \textit{double-spending} problem: in order to be validated, each input transaction is checked if used or unused. Only unused transactions are accepted.

\subsection{A probabilistic consensus: Proof-of-Work} \label{s:bitcoin-pow}
\begin{figure}[pht]
	\begin{center}
		\includegraphics[width=9cm]{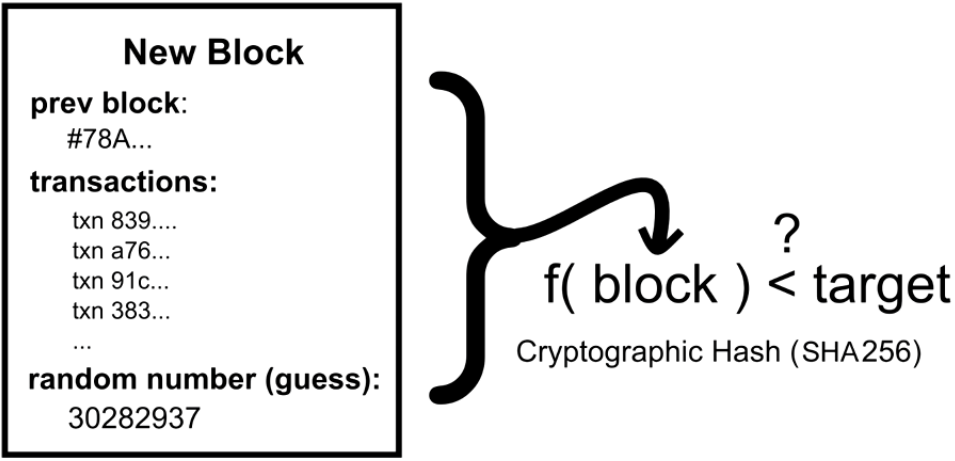}
	\end{center}
	\itcaption{Proof-of-Work as a computational puzzle to solve a block}
	\label{fig:blockchain-block}
\end{figure}

To achieve consensus on transactions order over the blockchain and to prevent duble-spending, Bitcoin's network is based on the \textit{Proof-of-Work} (PoW) algorithm. It consists in a computational intensive hashing task. Specifically, to solve a block miners need to find a random number (also called \emph{guess}) which acts as a proof of work (see Figure \ref{fig:blockchain-block}); indeed, such a number concatenated to transactions collected inside the block has to provide the overall hash of the block less than the target number. The target number is regulated according to the so-called \textit{blockchain difficulty} that regulates the average time spent by miners to accomplish such a task and create a new block.

Once a miner solves a block, it broadcasts that block over the network. They consider such a block as the latest of the chain and start mining new blocks to be appended on it. For the sake of simplicity, we can say that once a miner has created a new block, it becomes part of the chain. However if multiple miners concurrently add a block, a transient fork is created which is usually resolved over time by concatenating other blocks as by design miners always consider the longest chain. Miners are incentivised to correctly support the network due to rewards obtained for successfully mining blocks and for fees that each transaction can optionally acknowledge to them \cite{bitcoin2008, incentive}. 
\smallskip

When a block is part of the chain, it means that all miners have agreed on its content, hence it is practically non-repudiable and persistent unless an attacker has the majority of miners' hash power. In such case the adversary is able to create a chain fork or reverse a transaction. 
Assuming a majority of hash power controlled by honest miners, the probability of fork, with depth $n$, is $\mathcal{O}(2^{-n})$~\cite{bonneau2015sok}. 
Since the probability of forks decreases exponentially, the users simply waiting for a small number of blocks to be appended (e.g. 6 blocks in Bitcoin), can be sure that their transactions are permanently included with high confidence (avoiding the double spending attack).
However, in \cite{eyal2014majority} the authors showed as "\textit{majority is not enough: Bitcoin mining is vulnerable}" if only $25\%$ of the computing power is controlled by an adversary.

Although such systems provide strong integrity properties, PoW-based blockchains have a main drawback: \textit{performance}. This lack of performance is mainly due to the broadcasting latency of blocks on the network and to the time-intensive task of PoW. Indeed thousands of miners spread world-wide are required to render tampering with transactions computational infeasible and broadcast blocks over this kind of network topology takes very long.
Thus as a matter of fact, each transaction stored on a blockchain has a high confirmation time, which causes an extremely low transaction throughput. In Bitcoin, the average latency is 10 minutes, and the throughput is about 7 transactions per second, while for Ethereum the latency is on average 10 seconds with 15 transactions per second as throughput \cite{bonneau2015sok}. Moreover, PoW is energetically inefficient leading a huge waste of money and resources to make the hashing computations.

\section{Ethereum} \label{s:ethereum}
Ethereum \cite{wood2014ethereum} is the second main project (open source) of blockchain, following Bitcoin. It builds on the idea of \textit{smart contracts}, immutable programs deployed and executed autonomously on blockchain as transactions. It is featured by a Turing-complete programming language which introduce to the distributed computation system that allows anyone to write smart contract and therefore decentralised applications. This widens the application domains of blockchain making it a computational infrastructure with any of the participant in control of it.

Like Bitcoin, the shared state is managed by enrolled accounts which are featured by a balance of the Ethereum's cryptocurrency, i.e. \textit{ether}. The ether is the main internal crypto-fuel of Ethereum, and is used to pay transaction's fees. Like Bitcoin, nodes run a proof-of-work consensus to achieve agreement on transactions order and to maintain the shared state consistent. Every time a transaction is accepted, the state is updated by the Ethereum State Transition Function.

There are two types of accounts: \textit{externally owned accounts}, controlled by private keys, and \textit{contract accounts}, controlled by their contract code. The first are similar to the classic accounts of Bitcoin, whereas the others are allowed to allowed to read/write internal storage interacting with smart contracts.

\subsection*{Messages and Transactions}
Transactions, which in Ethereum are sent from external accounts, are quite similar to the Bitcoin's one. Two more fields have been introduced:
\enu
\item \textbf{Gas Price}: a scalar value representing the number of Ether to be paid for each computational step of the transaction?s execution;
\item \textbf{Gas Limit}: a scalar value representing the maximum amount of gas that should be used in executing the transaction.
\une

In order to prevent accidental or hostile infinite loops or other computational wastage in code, each transaction must have a computation limit on code execution. The unit \textit{gas} setup a fee system in such a way is required by attackers to pay, proportionately to every resource that they consume, including computation, bandwidth and storage; hence, any transaction that leads to the network consuming a greater amount of any of these resources must have a gas fee roughly proportional to the increment.

Moreover, Ethereum introduce the concept of \textit{message}. Messages are produced by contracts and essentially represent the call to a smart contract function. These are virtual objects that are never serialised, they exist only in the Ethereum execution environment and are sent by contract accounts. A message contain: 
\ite
\item the sender of the message;
\item the recipient of the message;
\item the amount of ether to transfer alongside the message;
\item the Gas Price.
\eti
\subsection*{GHOST Protocol}
Blockchains with fast confirmation times currently suffer of low security due to a high stale rate. Blocks take a certain time to propagate through the network, e.g. if miner A mines a block and then miner B happens to mine another block before miner A's block propagates to B, miner B's block will end up wasted and will not contribute to network security by violating consistency: forks \cite{eth-white}. Moreover blockchains which quickly produce blocks are very likely to lead to one mining pool, i.e. batch of miners working together to get higher computational power, having a large enough percentage of the network hashpower to have de facto control over the mining process.

In Bitcoin forks are solved by converging to the longest chain of blocks, indeed Ethereum applies the "Greedy Heaviest Observed Subtree" (GHOST) protocol \cite{SompolinskyZ15}. Here to calculate the main chain, also the stale descendants of the block's ancestor (in Ethereum jargon, i.e. \textit{uncles}) are added to the calculation on which block has the largest total proof of work backing it.

\begin{figure}[pht]
	\begin{center}
		\includegraphics[width=11cm]{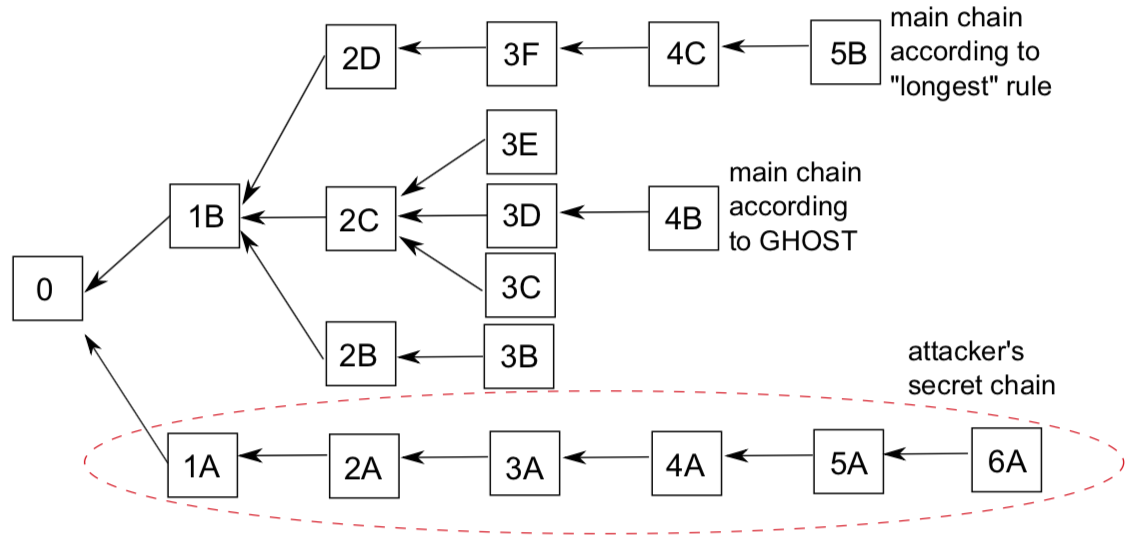}
	\end{center}
	\itcaption{Block tree where the longest chain is not the one selected by the GHOST protocol.}
	\label{fig:ghost}
\end{figure}

\figref{fig:ghost} illustrates a chain selected by the GHOST protocol. Consider the block 1B, it is supported by blocks 2B, 2C, and 2D that extend it directly, and include it in their chain. Similarly, blocks 3C, 3D, and 3E support both 1B and 2C as part of their chain. At each fork is selected the block leading to the heaviest subtree as part of the main chain. In such a way every addition of new forks to a subtree, makes harder the omission of the root from the main chain. Moreover to address the security issue of centralisation, this protocol also provide block rewards to stales: a stale block receives 87.5\% of its base reward, and the nephew that includes the stale block receives the remaining 12.5\%. 

Ethereum implements a simplified version of GHOST which only goes down seven levels. Specifically, a stale block can only be included as an uncle by the 2nd to 7th generation child of its parent, and not any block with a more distant relation. Calculations show that five-level GHOST with incentivisation is over 95\% efficient even with a 12s block time, that is the time between the creation of two blocks in Ethereum compared to 10 minutes in Bitcoin.

\subsection{Toward alternative consensus schemes: Proof-of-Stake} \label{s:ethereum-pos}
Ethash is the PoW implementation for Ethereum. Like in Bitcoin it suffers of the main drawbacks of probabilistic algorithms. Thus, has been proposed by Ethereum the \textit{Proof-of-Stake} (PoS), with the Nxt cryptocurrency \cite{nxt}, as alternative consensus algorithm for public blockchain. PoS employs a deterministic (pseudo-random) solution to define the creator of the next block and the chance that an account is chosen depends on its wealth (i.e. the stake). In PoS the blocks are said to be \textit{minted}, rather than mined. Moreover, generally all coins are created at the beginning, and thus the total number of coins never change afterwards, but there exist some PoS versions where new coins can be created. Therefore, in the original version of the PoS there are not any block rewards, hence the minter take only the transaction fees \cite{popov2016probabilistic}.

In the Ethereum's PoS implementation, called Casper \cite{eth-pos}, a set of validators take turns proposing and voting on the next block. The weight of each validator's vote depends on the size of its ether deposit (i.e. the stake). The process of creating and agreeing to new blocks is then done through a consensus algorithm that all current validators can participate in. Two major types of consensus can be identified for PoS protocol:
\enu
\item \textbf{Chain-based PoS}: the algorithm pseudo-randomly selects a validator during each time slot (eg. every period of 10 seconds might be a time slot) to propose a block. The block is then appended to the blockchain 
\item \textbf{BFT-style PoS}:  validators are randomly assigned the right to propose blocks, but agreeing on which block is canonical is done through a multi-round process where each validator sends a "vote" for some specific block during each round, and at the end of the process all (honest and online) validators permanently agree on whether or not any given block is part of the chain. Note that blocks may still be chained together; the key difference is that consensus on a block can come within one block, and does not depend on the length or size of the chain after it.
\une

Differently from PoW, the PoS algorithms do not require waste of energy because they do not need long computational periods. Hence the minting process is performed by coin owners, putting money (i.e. a deposit) at stake. Here honest validators receive small rewards for making blocks but, security against malicious behaviours should be granted by penalties. Indeed in this case, the costs of reverting transactions are hundreds or thousands of times larger than the rewards that they got in the meantime. As the Ethereum's creator, Vitalik Buterin, wrote in a blogpost \cite{eth-pos-medium} :"In Proof-of-Stake, security comes not from burning energy, but rather security comes from putting up economic value-at-loss".

However many chain-based PoS algorithms are rising with a reward policy for producing blocks without penalties \cite{peercoin} and this lead to the \textit{nothing~at~stake} problem. Thus this issue, in case of multiple chains, incentivise validators to sign blocks on every chain since it does not cost anything to them. In this scenario a validator is encouraged to act badly and creates as many blocks as possible leading to multiple forks and consequently to double spend attacks.

\subsection{Ethereum private networks} \label{s:ethereum-private}
There exist different implementations of the Ethereum protocol, most of them offer the possibility to be used in consortium chains for private settings. Two of the most interesting Ethereum's client are Geth \cite{geth}, the Ethereum implementation in Golang language, and Parity \cite{parity}, a Rust-based implementation. Both of them offer, through special configurations, the possibility of build permissioned private blockchain environments with lightest consensus protocols.

These type of chains are used in development testnets where distributed applications are tested and debugged before the deployment on the main Ethereum blockchain. Moreover private networks, optimised for permissioned settings, can be also used for enterprise use cases where performance and privacy are important.
\smallskip

Beyond the currently PoW consensus, BFT-based algorithms can be employed in such a scenario to guarantee highest performances. In the next chapter we will see which are the most prominent algorithms and some of their use cases.

\section{Hyperledger} \label{s:hyperledger}
The Hyperledger consortium is an ombrella project under the Linux Foundation that incubate a large number of blockchain projects. Fabric \cite{fabric} is the first released smart contract, here called \textit{chaincode}, blockchain system for permissioned settings. It offers a customisable platform in terms of pluggable consensus mechanism, regulation on member enrolment and visibility of data. Avoiding the use of proof-of-work prevents Fabric to be used in trust-less settings, but its performance is remarkable and enabled a wide range of industrial applications.

Firstly released under the most conform characteristics of a state-machine replication system, i.e. release \textit{v0.6} \cite{Cachin2016ArchitectureOT}, actually a more elaborate design has been adopted. The last release \textit{Fabric v1.0} \cite{fabricv1} separates the execution of chaincode transactions from the ordering to ensure agreement and consistency. This new architecture promises to bring better scalability. The fundamental key features of this architecture are:
\ite
\item Channels for sharing confidential informations;
\item An Ordering Service to guarantee transactions agreement and consistency, over the network;
\item An Endorsement policy for transactions;
\item CouchDB world state which supports queries;
\item A customisable Membership Service Provider (MSP).
\eti

At the beginning, the first release of Fabric had come with a native implementation of PBFT \cite{castro1999practical} as consensus protocol. With the new version, the Ordering Service can be provided by the Apache Kafka protocol \cite{kafka}, i.e. a crash tolerant pub/sub consensus and, by a noops protocol, called SOLO, which essentially delegate the ordering to one peer. Furthermore a new implementation of PBFT is under development.

\chapter{Distributed Consensus}\markchapter[Chapter~\thechapter:~Distributed Consensus]
\label{c02:consensus}
% !TEX root = ../../Tesi-1side.tex
Modern systems, evolved from centralised to distributed architectures where interconnected computers, i.e. \textit{replicas}, communicate and collaborate for a common goal combining their resources, i.e. \textit{Distributed Computing}. Replication in computing helps to improve availability, reliability and fault-tolerance. There exists two types of replication: (i)~\textit{data~replication}, used in Distributed Database systems where data are stored across replicas, (ii)~\textit{computational~replication} where replicas execute the same computing task many times. Furthermore replicated systems can be identified as:
\ite
\item \textbi{active replication}: replicas process the same request;
\item \textbi{passive replication}: each single request is processed on a single replica and the result is dispatched to the other replicas.
\eti

The simplest active replication scheme is the \textit{primary/backup} (or master-slave in the database field) where a primary replica is in charge to order all the requests and dispatch them to all the backups. On the other side, if any replica processes a request and then distributes a new state, then this is a \textit{multi-primary} scheme (or multi-master).

A fundamental problem in distributed computing is coordinating replicas operations also in unreliable settings, i.e. \textit{Distributed Consensus}. The most common paradigm for implementing this type of services is the State Machine Replication (SMR) \cite{schneider1990implementing}, wherein a deterministic state machine is replicated across a set of nodes, such that it functions as a single state machine despite failures. Replicas start in the same initial state and apply operations in the same order. Considering the client/server paradigm, SMR systems ensure resiliency against single-point-of-failure, contrarily to services using a single centralised server. Clients submit operations to the replicated server system, i.e. \textit{transactions}, each one causing a state transition and return a deterministic result. 

Blockchain is an implementation of a distributed database based on active replication with a multi-primary scheme.

To be feasible for an SMR system, distributed consensus algorithms must guarantee two main properties:

\begin{enumerate}
	\labitem{$\property{}$}{prop:1} \textit{\textbf{Safety.} This property states that the algorithm should do anything wrong during its normal execution. This property prevent unwanted executions and ensure different properties like consistency, validity agreement, and so on. If it is violated at some point in time, safety property will be never satisfied again.}
	
	\labitem{$\property{}$}{prop:2} \textit{\textbf{Liveness.} This property ensures that eventually something good happens. It is a property of a distributed system which grant that the algorithm works properly in time and that sooner or later every execution completes correctly.}
\end{enumerate}

According to these properties different consensus algorithms have been proposed in the last years. All of this protocols rely on different network models and guarantee correctness despite some fault model. The fault model and the synchrony assumptions play a fundamental role for safety and liveness guarantees in SMR systems. Fixing a particular type of fault and the assumptions on network synchrony, one can characterise the model by its resiliency, i.e. the maximum number of faults which can be tolerated in any protocol in the given model.

\section{Network models}\label{s:consensus-network}
Typically replication protocols, to guarantee resiliency, require an upper bound on the time for messages to be delivered and on processor clocks synchronisation. In \textit{synchronous} environments these upper bound are fixed and defined a priori, therefore achieving consensus results easy. However synchronous solutions are unfeasible in real environments where networks can be partitioned, messages may be delayed or lost, nodes may crash, may be subverted by an attacker and other unpredictable events. 

\textit{Asynchronous} systems avoid the necessity of any assumption about synchronised clocks and timely network behaviours. Therefore protocols designed in this scenario could provide the best resiliency to any type of fault. However due to a fundamental discovery by Fischer et al., i.e. the celebrated "FLP impossibility result" \cite{fischer1985impossibility}, which shows that the existence of timing bounds is necessary to achieve any resiliency, is proved that in this model, if even a single processes can crash, no consensus can be achieved.

The most accepted network model relies on \textit{eventual~synchrony} \cite{dwork1988consensus}. This simulates an asynchronous network which eventually, after undefined bound in time, behave synchronously and delivers all the messages. Protocols in this model never violate safety as long as the trust assumptions on the kind and the number of faulty nodes are met. During the synchronous period, the protocol terminate correctly, guaranteeing liveness.

\section{Failure Classification}\label{s:consensus-fail-model}
Distributed systems can be subject to several issues caused by replication like random communication delays and systems failures. These brings to the systems problems in distributed consensus, and may cause the violation of safety and liveness properties. 

In \cite{CRISTIAN1995158}, Cristian et al. propose a classification of failures in several nested classes, so that protocol's complexity can be measured with the size of the class of failures it tolerates. Failures are classified with respect to system components: a failure occurs when a component does not behave in the correct manner.

An \textit{omission} failure occurs when a component never gives the expected output to a specified input. A \textit{timing} failure occurs when the component responds too early, too late or never to a request. A \textit{Byzantine} failure \cite{lamport1982byzantine} occurs when the component does not behave correctly. More precisely a Byzantine fault can be cause by an omission fault, a timing fault or if the component gives different output from the one specified. An important subclass of this failure is the so called \textit{authentication-detectable} Byzantine failure, where any corruption of messages is detectable by using a message authentication protocol, e.g. public-key cryptosystems based on digital signatures.

\begin{figure}[pht]
	\begin{center}
		\includegraphics[width=12cm]{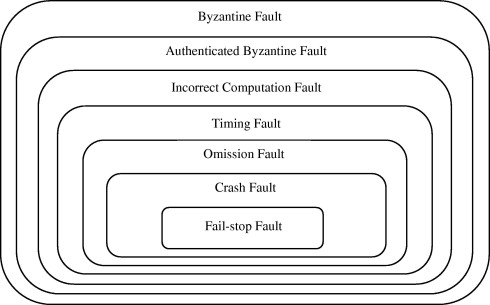}
	\end{center}
	\itcaption{Byzantine failures model.}
	\label{fig:byz-model}
\end{figure}

\smallskip
\figref{fig:byz-model} shows the Byzantine Failures model diagram where is explained the classification of failures. Crash failures, i.e. when a component stops to work, are a subclass of omission failures (\textit{Fail-stop} failure is a special class of crash, where after a first omission a component omits to respond to all subsequent input events). Omission failures are a subclass of timing failures, timing failures are a subclass of authenticated-detectable Byzantine failures which are a subclass of the class of all possible failures, the Byzantine failures.

The nested nature of the failure classes makes it easy to compare fault-tolerant protocols.

\section{Atomic Broadcast}\label{s:consensus-atomic-bcast}
The most relevant algorithm for SMR systems, for guaranteeing agreement and order on requests, is the \textit{atomic~broadcast} (or \textit{total order broadcast}) consensus.	

Atomic broadcast is a \textit{reliable broadcast} which ensures also that correct nodes output or \textit{deliver} the same sequence of messages, i.e. total order. Specifically it ensures the following safety properties \cite{hadzilacos1993fault}:

\begin{enumerate}
	\labitem{$\property{}$}{prop:3} \textit{\textbf{Validity.} If a correct node $p$ broadcasts a message $m$, then $p$ eventually delivers $m$.}
	
	\labitem{$\property{}$}{prop:4} \textit{\textbf{Agreement.} If a message $m$ is delivered by some correct node, then $m$ is eventually delivered by every correct node.}
	
	\labitem{$\property{}$}{prop:5} \textit{\textbf{Integrity.} No correct node delivers the same message more than once; moreover, if a correct node delivers a message $m$ and the sender $p$ of $m$ is correct, then $m$ was previously broadcast by $p$.}
	
	\labitem{$\property{}$}{prop:6} \textit{\textbf{Total Order.} For messages $m_1$ and $m_2$, suppose $p$ and $q$ are two correct nodes that deliver $m_1$ and $m_2$. Then $p$ delivers $m_1$ before $m_2$ if and only if $q$ delivers $m_1$ before $m_2$.}	
\end{enumerate}	

\smallskip
There exists several implementations of the atomic broadcast consensus, however a fair comparison between each algorithm is still considered challenging \cite{cachin2017blockchains} as some of them require a \textit{synchronous} network, while others can work in a \textit{eventually synchronous} network which may strongly impact performance and scalability. Therefore safety and liveness properties in each model are guaranteed under different trust assumptions. Typically the generic trust assumption for a system with $n$ nodes says that no more than $f < n/k$ nodes became faulty, thus the other $n-f$ nodes are correct. For instance, in SMR systems considering the simplest replication method, i.e. the primary/backup, and the special fail-stop model, only $f+1$ replicas are required by tolerating up to $f$ faulty nodes.

\begin{table}[!htb]
\centering
\begin{tabular}{c|c|c|c|}
\cline{2-4}
\begin{tabular}[c]{@{}c@{}}Failure\\ Mode\end{tabular} & Synchronous & Asynchronous & \begin{tabular}[c]{@{}c@{}}Partially\\ Synchronous\end{tabular} \\ \hline
\multicolumn{1}{|c|}{Fail-Stop} & $f$ & $\infty$ & $2f+1$ \\ \hline
\multicolumn{1}{|c|}{Omission} & $f$ & $\infty$ & $2f+1$ \\ \hline
\multicolumn{1}{|c|}{\begin{tabular}[c]{@{}c@{}}Byzantine\\ Authentication\end{tabular}} & $f$ & $\infty$ & $3f+1$ \\ \hline
\multicolumn{1}{|c|}{Byzantine} & $3f+1$ & $\infty$ & $3f+1$ \\ \hline
\end{tabular}
\caption{Smallest number of nodes for which a f-resilient consensus protocol exists.}
\label{table:fault-cons}
\end{table}

In \tabref{table:fault-cons} is shown the comparison of trust assumptions required for consensus, considering all the failure models introduced before in different synchrony models.

All those algorithms can be categorised in two family: (i) \textit{Crash Tolerant} and (ii) \textit{Byzantine Fault Tolerant} (BFT).

The main references for those families are (i) Paxos \cite{lamport1998part, lamport2001paxos}, proposed by Lamport in its original version only tolerant to a crash of $t<n/2$ nodes, and (ii) Practical Byzantine Fault Tolerance (PBFT), proposed by Castro and Liskov \cite{castro1999practical}, which is able to resist to $f<n/3$ Byzantine nodes.
% Si Potrebbe aggiungere una descrizione più approfondita dei due. Per PBFT interessante quanto scritto sulla tesi di Tendermint

The PoW algorithm, even though is not related to neither families, can be categorised as a BFT algorithm on a SMR model. Rigorously analysed in the technical report \cite{MillerTechRep}, it is modelled in a network with a near-infinite number of nodes, with each node representing a very small unit of computing power and having a very small probability of being able to create a block in a given period. In this model, the protocol has 50\% fault-tolerance assuming zero network latency (synchronous network), $\sim$46\% (Ethereum) and $\sim$49.5\% (Bitcoin) fault-tolerance under actually observed conditions, but goes down to 33\% if network latency is equal to the block time, and reduces to zero as network latency approaches infinity (asynchronous network) \cite{eth-pos}.

\section{Consensus in Permissioned Blockchain}\label{s:consensus-perm-block}
Blockchain is essentially a distributed database holding a list of records controlled by a trustless p2p network. Actually the modern distributed database implementations, e.g. DynamoDB \cite{DeCandia-dynamo}, Cassandra \cite{cass-db}, Facebook TAO \cite{TAOFace}, referencing the CAP Theorem, favour availability and partition tolerance over consistency. Furthermore these systems offer the concept of \textit{eventual consistency} where replicas in a distributed database eventually converge to identical copies. Such scenarios address the scalability and availability issues associated with strong consistency semantics, i.e. linearizability \cite{HerlihyLinear, Vukolic16}.

\paragraph{CAP Theorem} The CAP Theorem~\cite{brewer2012cap} states that in a distributed data store only two out of the three following properties can be ensured: \textit{Consistency} (C), \textit{Availability} (A) and \textit{Partition Tolerance} (P). Thus any distributed data store can be characterised on the basis of the (at most) two properties it can guarantee, either CA, CP or AP. However, since replicated systems may be subject to failures that can be thought as partition, partition-tolerance must be guaranteed, thus only CP or AP systems are feasible. 

\bigskip
Even if the blockchain system is a distributed database, it should enforce strong consistency, i.e. atomic broadcast, among transactions, to prevent issues such as the double-spending attack. However PoW-based permissionless blockchain cannot guarantee strong consistency because they run a sort of \textit{eventual consensus} where forks are possible and eventually resolved, i.e. such as the longest branch rule of Bitcoin or the GHOST protocol in Ethereum. These systems do not guarantee atomic broadcast and cannot be identified as SMR \cite{Vukolic16}. The eventual consensus in PoW (and PoS) leads to favour availability over consistency guaranteeing only eventual consistency \cite{eth-pos}.

\smallskip
Conversely to permissionless blockchains, permissioned blockchains usually do not employ cryptocurrencies and their purpose is more oriented to exchange of asset and transactions. They can be faster by design, implementing a classical SMR schema. Here the authentication of nodes changes the trust level, allowing thus to employ a consensus schema lighter than the PoW.  A report of Swanson compares the two models \cite{swanson2015consensus}. 

Hence blockchain are turning back to classical, strongly-consistent BFT distributed consensus, based on atomic broadcast. These systems avoid the possibility of forks and, in blockchain jargon this property is identified as \textit{consensus finality} introduced in \cite{vukolic2015quest} (in Appendix \ref{a:finality} is detailed the proof of strong consistency in finality).

\begin{enumerate}
	\labitem{$\property{}$}{prop:7} \textit{\textbf{Finality.} Given two correct nodes $n_1$, $n_2$, if $n_1$ appends a block $b$ to its copy of the blockchain before appending block $b'$, with $b \neq b'$ then no correct node $n_2$ appends block $b'$ before $b$ to its copy of the blockchain.}
\end{enumerate}

Classical BFT-based consensus algorithms are strongly consistent, e.g. PBFT, and guarantee finality to blockchains based on it. The most feasible algorithms for permissioned blockchain belongs essentially to two implementations:
\begin{itemize}
\item \textit{Mining Rotation algorithm} \cite{publicVSprivate1}. Distinguished nodes are in charge to seal all of the blocks of the blockchain electing a leader periodically.
\item \textit{PBFT (Practical Byzantine Fault Tolerant)} \cite{castro1999practical}. Traditional Byzantine Fault Tolerant consensus or similar.
\end{itemize}

This trend of moving on strongly-consistent BFT-based consensus in blockchain is exemplified by all the emerging practical systems which are rising. The most important blockchain technologies that offer BFT-based distributed ledger systems are: (i) Hyperledger Fabric with Apache Kafka \cite{kafka} (i.e. a crash tolerant pub/sub) and with PBFT, (ii) Tendermint \cite{kwon2014tendermint} a BFT consensus protocol for blockcahin similar to PBFT, (iii) R3 Corda with Raft \cite{ongaro2014search} (i.e. a popular variant of Paxos, thus crash tolerant) and BFT-SMaRT \cite{bftsmart} (i.e. the most advanced and tested implementation of a BFT consensus), (iv) Multichain \cite{multichain}, (v) Sawtooth Lake \cite{sawtooth} with the Proof-of-Elapsed-Time (a.k.a. PoET, i.e. a novel protocol based on the insight that PoW essentially imposes a mandatory but random waiting time for leader election) and many others. An exhaustive comparison between different algorithms employed by existing technologies among those properties is available in \cite{cachin2017blockchains}. The work compares this technologies and their requirements in the sense of safety and liveness properties. However from this study emerge a set of challenges.

\smallskip
BFT-based protocols proposed for permissioned blockchain guarantee strong consistency but have to give up availability or partition-tolerance because the CAP Theorem. Moreover a fair comparison between each algorithm is still considered challenging \cite{cachin2017blockchains} as some of them requires a \textit{synchronous} network, while others can work in a \textit{eventually synchronous} network which may strongly impact on performance and scalability. Indeed, this protocols are based on developers ideas, not relying on established knowledge on blockchain. Most of them come without formal expression of their trust assumptions and security model. There is not an agreed consensus in the industry and this generate a lot of confusion in the public opinion \cite{cachin2017blockchains}.

\smallskip
In the next section we propose an analysis of two Mining Rotation algorithms against the PBFT one, through the help of the CAP Theorem, to advocate liveness and safety guarantees.

\chapter{Assessing Proof-of-Authority Consensus for Permissioned Blockchain through a CAP Theorem-based Analysis}\markchapter[Chapter~\thechapter:~Assessing Proof-of-Authority Consensus for Permissioned Blockchain through a CAP Theorem-based Analysis]
\label{c10:poa}
% !TEX root = ../../Tesi-1side.tex

Recently, a new line of research on \textit{distributed consensus} is arising, as it plays a key role in the blockhain ecosystem. In permissioned settings, dominant candidates are \textit{Byzantine fault tolerant} (BFT) algorithms such as the Practical BFT (PBFT). Indeed, BFT-like algorithms have been widely investigated for permissioned blockchains with the aim of outperforming PoW while ensuring adequate fault tolerance. 

%Proof-of-Authority
Proof-of-Authority (PoA) \cite{poa} is a new family of BFT algorithms which has recently drawn attention due to the offered performance and toleration to faults. It is currently used by Parity~\cite{parity} and Geth~\cite{geth}, two well-recognised clients for permissioned setting of Ethereum. Intuitively, the algorithms operate in rounds during which an elected party acts as \emph{mining leader}~\cite{publicVSprivate1} and is in charge of proposing new blocks on which distributed consensus is achieved. Differently from PBFT, PoA requires less message exchanges hence provides better performance~\cite{dinh2017blockbench}. However, the actual consequences of such performance improvement is quite blurry, especially in terms of availability and consistency guarantees in a realistic \emph{eventually synchronous} network model such as the Internet~\cite{cachin2011introduction}). 

%Analysis approach for Aura
To this aim, we take into account two of the main PoA implementations, named Aura~\cite{aura} and Clique~\cite{clique}, which are used by Ethereum clients for permissioned-oriented deployments. In particular, the lack of appropriate documentation and analysis prevents from a cautious choice of PoA implementations with respect to provided guarantees, fault tolerance and network models. 

\smallskip
In this chapter, we first derive the actual functioning of the two PoA algorithms, both from the scarse documentation and directly from the source code. Then, we conduct a comparison with PBFT in terms of security and performances. Specifically, we conduct a qualitative analysis in terms of the CAP theorem, i.e., by assessing which "\textit{two out of three}" properties between \textit{consistency}, \textit{availability} and \textit{partition tolerance} can be assured at the same time. Finally, we assess performances in terms of message exchange. The analysis assumes an \textit{eventually synchronous network} and the presence of Byzantine nodes.
%
%Results
The conducted analysis results that PoA algorithms favour availability over consistency, oppositely to what PBFT guarantees. 
In terms of latency, measured as the number of message rounds required to commit a block, PBFT lies in between Aura and Clique, outperforming the former and being worse than the latter.
These results suggest that PoA algorithms are not actually suitable for permissioned blockchains deployed over the Internet, because they do not ensure consistency, and strong data integrity guarantees are usually the reason why blockchain-based solutions are employed. We advocate that PBFT is a better choice in this case, although its performance can be worse than some PoA implementations, thus their application may highly depend from the scenario.

\smallskip
\noindent\textit{Contributions.} The content of this chapter has been published at the conference ITASEC18, the contributions can be summarised as follow:

\begin{itemize}
	\item we detailed two of the most prominence version of PoA algorithms by reversing source code and online documentation;
	\item we proposed an approach to evaluate consensus algorithms through the CAP;
	\item we conducted such an analysis on the two PoA algorithms as well as on PBFT;
	\item we analysed performances of PoA and PBFT.
\end{itemize}

\smallskip
\noindent\textit{Chapter Structure.} Section \ref{s:poa-related} introduces background concepts and comments the closest related work. Section \ref{s:poa} introduces the PoA consensus schema. Section \ref{s:poaVSpbft} analyses  PoA algorithms with respect to ensured guarantees and performance. Finally, Section \ref{s:poa-discussion} concludes the work.

% !TEX root = ../../Tesi-1side.tex

\section{Related Work} \label{s:poa-related}
The most prominent consensus schema for consensus in distributed computing is so-called Byzantine Fault Tolerant (BFT). Protocols of such type are able to tolerate arbitrarily subverted nodes trying to hinder the achievement of an consistent agreement as we detailed in Chapter \ref{c02:consensus}.

The \textit{Practical Byzantine Fault Tolerance} (PBFT)~\cite{castro1999practical} is one of the most well-established BFT algorithms. However many other BFT algorithms have been proposed, mainly for improving PBFT performance; among others we can cite \textit{Q/U}, \textit{HQ}, \textit{Zyzzyva}, \textit{Aardvark}, see the survey in~\cite{tseng2016recent} for further details of each solution. 

The wide interest on blockchain has prompted substantial research efforts on distributed consensus schema, specifically towards new ad-hoc BFT ones. 
In \cite{mattila2016blockchain} the author reviews well-known families of consensus algorithms for both permissionless and permissioned blockchains. This includes Proof-of-Work~(PoW), Prof-of-Stake~(PoS), Delegated Proof-of-Stake~(DPoS), Proof-of-Activity~(PoW/PoS-hybrid), Proof-of-Burn~(PoB), Proof-of-Validation~(PoV), Proof-of-Capacity~(PoC or Proof-of-Storage), Proof-of-Importance (PoI), Proof-of-Existence~(PoE), Proof-of Elapsed Time~(PoET), Ripple Consensus Protocol and Stellar Consensus Protocol~(SCP).
Although each algorithm is briefly described, they lack of any form of analysis in presence of Byzantine nodes under an eventual synchronous model. 

Understanding the most appropriate consensus algorithm among the plethora before is a challenging task that a few works have tried to tackle. For instance, Sankar et al. investigates in~\cite{sankar2017survey} the main differences between SCP and consensus algorithms employed in R3 Corda and Hyperledger Fabric. Similarly, Mingxiao et al. extensively compare in~\cite{mingxiaoreview} performances and security of PoW, PoS, DPoS, PBFT and Raft. While, Tuan et al. propose in~\cite{dinh2017blockbench} a practical benchmark for blockchain, named Blockbench, to systematically compare performances, scalability and security of multiple blockchain systems. 

From a more formal perspective, Vukoli\'{c} compares in~\cite{vukolic2015quest} PoW with BFT-like approaches introducing the distinguishing property of \textit{consensus finality}: the impossibility of reaching consensus without fully distributed agreement. In blockchain's jargon, it amounts to the impossibility of having forks. As expected, PoW does not enjoy consensus finality (as forks can happen), while all BFT-like approaches does (all parties reach agreement before consensus). 

More related to permissioned blockchain, Cachin and Vukoli\'{c} propose in~\cite{cachin2017blockchains} a thorough analysis of most-known permissioned systems and their underlying consensus algorithms in term of safety and liveness guarantees under eventual synchrony assumption. This work introduces for the first time a structured comparison among consensus algorithms, but it overlooks consistency and availability guarantees ensured by their usage; most of all it does not address PoA.

To sum up, none of the aforementioned works discuss the PoA consensus family. To the best of our knowledge, we believe this is the first work tackling the analysis of blockchain consensus algorithms from the perspective of the CAP theorem.
% !TEX root = ../../Tesi-1side.tex

\section{Proof-of-Authority Consensus} \label{s:poa}

\textit{Proof of Authority} (PoA) is a family of consensus algorithms for permissioned blockchain whose prominence is due to performance increases with respect to typical BFT algorithms; this results from lighter message exchanges. PoA was originally proposed as part of the Ethereum ecosystem for private networks and implemented into the clients \textit{Aura} and \textit{Clique}.

PoA algorithms rely on a set of $N$ trusted nodes called the \textit{authorities}. Each authority is identified by a unique \textit{id} and a majority of them is assumed honest, namely at least $N/2 + 1$. 
Consensus in PoA algorithms relies on a \textit{mining rotation} schema, a widely used approach to fairly distribute the responsibility of block creation among authorities~\cite{publicVSprivate1,gaetani2017blockchain,edcc17}. Time is divided into \textit{steps}, each of which has an authority elected as mining leader\footnote{For the cognitive, as there is no hash-based procedure like PoW, the consensus process is more appropriately called \emph{minting}. However, for the sake of presentation, we will continue using the wording mining.}.

\begin{figure*}[!thbp]
	\centering
	\subfigure[Aura]{
		\label{fig:aura}
		\includegraphics[height=4.5cm]{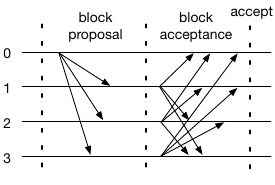}
	}\qquad
	\subfigure[Clique]{
		\label{fig:cliq}
		\includegraphics[height=4.5cm]{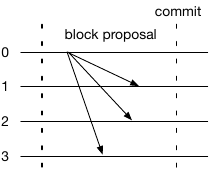}
	}
	\itcaption{Message exchanges of Aura and Clique PoA for each step. In this example there are 4 authorities with id 0,1,2,3. The leader of the step is the authority 0.}
	\label{fig:msg_ex1}
\end{figure*}

The two PoA implementations work quite differently: both have a first round where the new block is proposed by the current leader (\textit{block proposal}); then Aura requires a further round (\textit{block acceptance}), while Clique does not.
Figure~\ref{fig:msg_ex1} depicts the message patterns of Aura and Clique, which will be detailed in next subsections.

\subsection{Aura}\label{s:poa-aura}
\textit{Aura} (Authority Round) \cite{aura} is the PoA algorithm implemented in Parity~\cite{parity}, the Rust-based Ethereum client. 
The network is assumed to be synchronous and all authorities to be synchronised within the same UNIX time \textit{t}. The index $s$ of each step is deterministically computed by each authority as $s = t / step\_duration$, where $step\_duration$ is a constant determining the duration of a step. The leader of a step $s$ is the authority identified by the id $l = s\ mod\ N$.

Authorities maintain two queues locally, one for transactions $Q_{txn}$ and one for pending blocks $Q_{b}$.
Each issued transaction is collected by authorities in $Q_{txn}$. For each step, the leader $l$ includes the transactions in $Q_{txn}$ in a block $b$, and broadcasts it to the other authorities (\textit{block proposal} round in Figure~\ref{fig:aura}). Then each authority sends the received block to the others (round \textit{block acceptance}).
If it turns out that all the authorities received the same block $b$, they \textit{accept} $b$ by enqueuing it in $Q_{b}$. Any received block sent by an authority not expected to be the current leader is rejected. The leader is always expected to send a block, if no transaction is available then an empty block has to be sent.

If authorities do not agree on the proposed block during the block acceptance, a voting is triggered to decide whether the current leader is malicious and then kick it out. An authority can vote the current leader malicious because (i) it has not proposed any block, (ii) it has proposed more blocks than expected, or (iii) it has proposed different blocks to different authorities. 
The voting mechanism is realised through a \textit{smart contract}, and a majority of votes is required to actually remove the current leader $l$ from the set of legitimate authorities. When this happens, all the blocks in $Q_{b}$ proposed by $l$ are discarded. Note that leader misbehaviours can be caused by benign faults (e.g., network asynchrony, software crash) or Byzantine faults (e.g., the leader has been subverted and behaves maliciously on purpose).

\begin{figure*}[!thbp]
\centering
\subfigure[Step s1]{
\label{fig:fin_1}
\includegraphics[width=7.5cm]{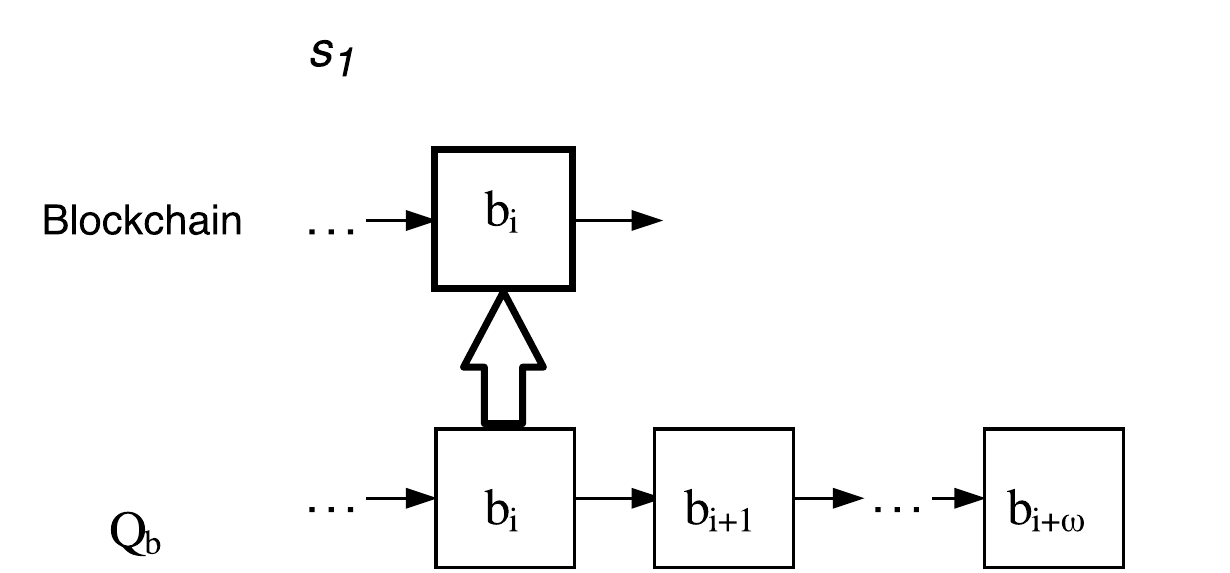}
}\qquad
\subfigure[Step s2]{
\label{fig:fin_2}
\includegraphics[width=7.5cm]{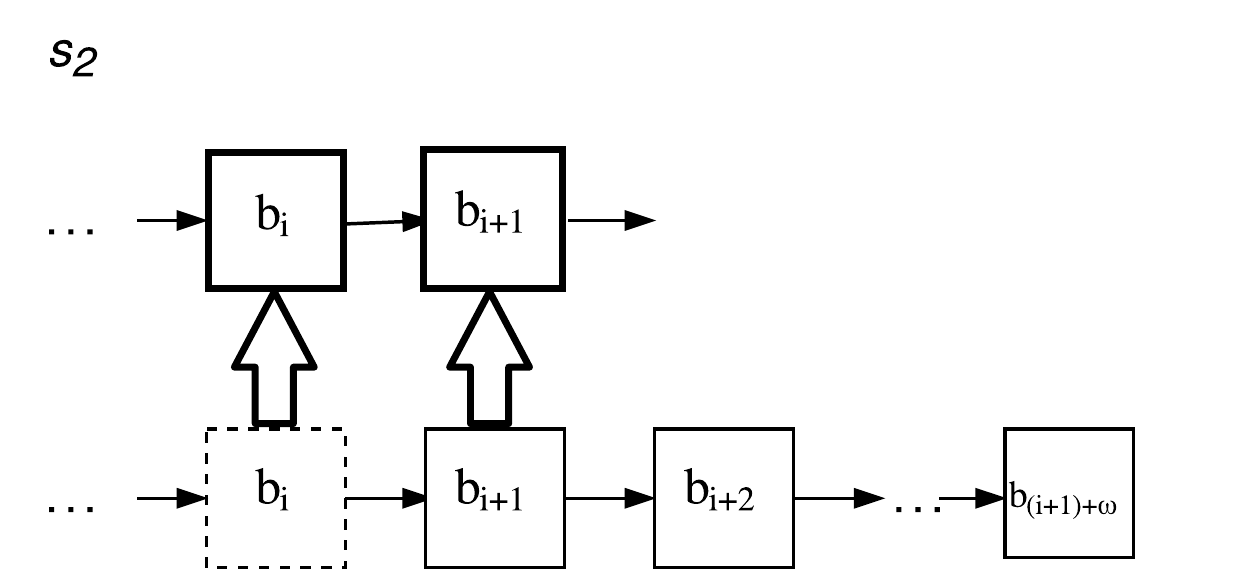}
}
\caption{Aura finality} 
\label{fig:aura_fin}
\end{figure*}

A block $b$ remains in $Q_{b}$ until a majority of authorities propose their blocks, then $b$ is committed to the blockchain. With a majority of honest authorities, this mechanism should prevent any minority of (even consecutive steps) Byzantine leaders to commit a block they have proposed, thus it guarantee consensus finality (\figref{fig:aura_fin}). Indeed any suspicious behaviour (e.g., a leader proposes different blocks to different authorities) triggers a voting where the honest majority can kick the current leader out, and the blocks they have proposed can be discarded before being committed.

\figref{fig:aura_fin} shows how Aura guarantees finality on two consecutive steps. At step $s_1$ on the blockchain are committed blocks till $b_i$, whereas blocks $b_{i+1} \cdots b_{i+\omega}$ are pending. The block $b_i$ can be committed, since $\omega = \frac{k}{2}+1$ further blocks have been proposed after $b_i$, and thus the block $b_i$ achieved finality. Likewise, at step $s_2$ the block $b_{i+1}$ achieves finality as the queue contains $\omega$ further blocks.

Summarising, the algorithm works as follows:
\begin{enumerate}
\item Nodes collects transactions in the transaction queue $Q_{txn}$;
\item For each step, the primary node seals and broadcasts a block to the network;
\item Nodes who receive a new block, verify its correctness and append it in the queue $Q_b$, waiting for finalisation;
\item When an epoch expires, each node examines the subset of all unfinalised blocks from $Q_b$. If a chain of waiting blocks has been finalised, than it is committed to the blockchain. Each committed transaction is removed from the queue $Q_{txn}$;
\end{enumerate}

\subsection{Clique}
\textit{Clique} \cite{clique} is the PoA algorithm implemented in Geth~\cite{geth}, the GoLang-based Ethereum client. The algorithm proceeds in \textit{epochs} which are identified by a prefixed sequence of committed blocks. When a new epoch starts, a special \textit{transition block} is broadcasted. It specifies the set of authorities (i.e., their ids) and can be used as snapshot of the current blockchain by new authorities needing to synchronise.

\begin{figure*}[!hbpt]
	\centering
	\subfigure[Time t1]{
		\label{fig:cliq_1}
		\includegraphics[width=6cm]{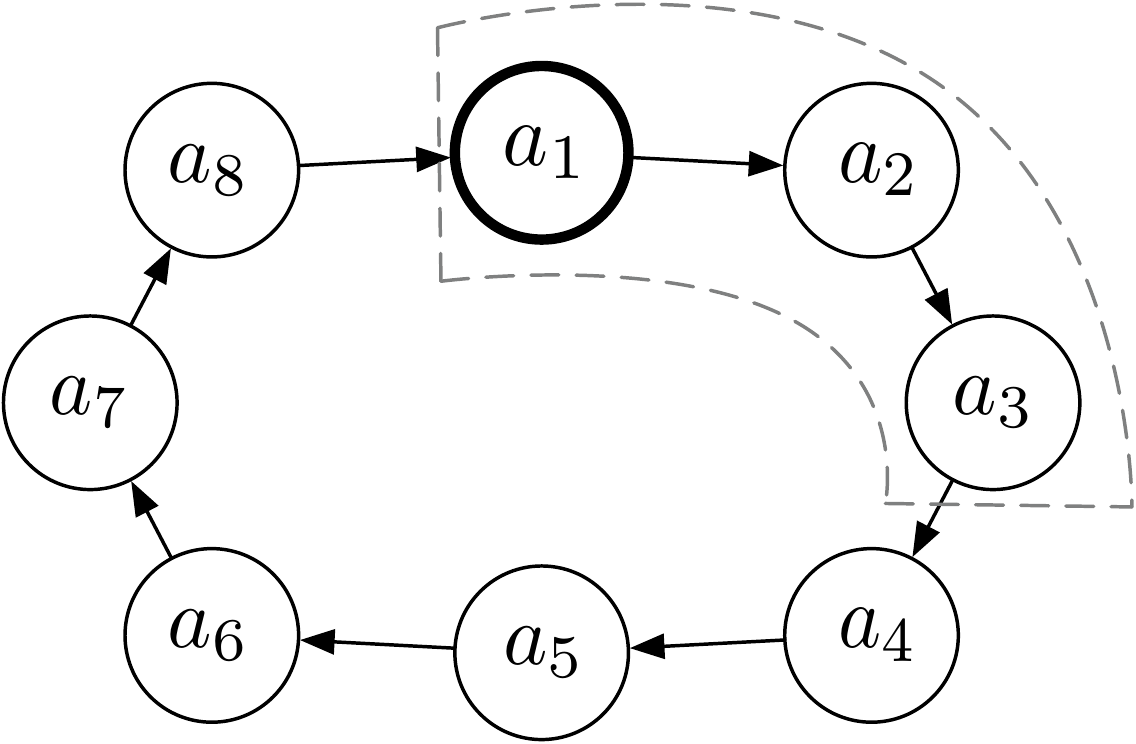}
	}\qquad
	\subfigure[Time t2]{
		\label{fig:cliq_2}
		\includegraphics[width=6cm]{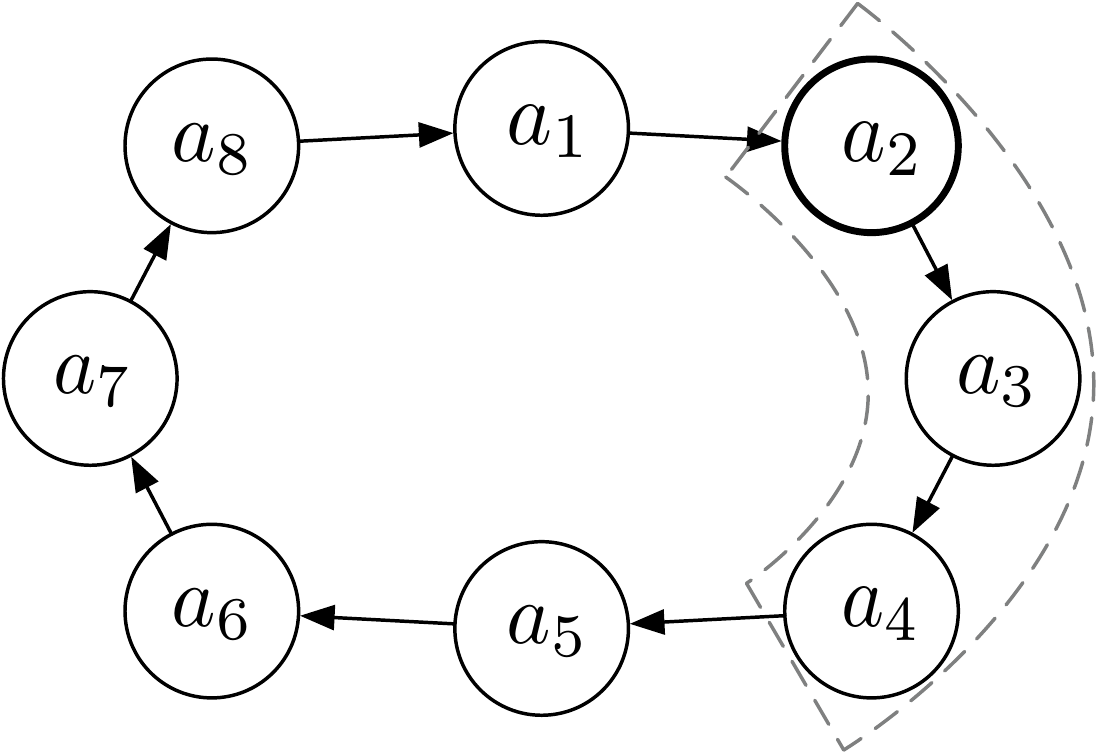}
	}
	\itcaption{Selection of authorities allowed to propose blocks in Clique.}
	\label{fig:cliq_rotation}
\end{figure*}

While Aura is based on UNIX time, Clique computes the current step and related leader using a formula that combines the block number and the number of authorities. Most of all, in addition to the current leader, other authorities are allowed to propose blocks in each step.
To avoid that a single Byzantine authority could wreak havoc the network by imposing a sheer number of blocks, each authority is only allowed to propose a block every $N/2 + 1$ blocks. Thus, at any point in time there are at most $N - (N/2 + 1)$ authorities allowed to propose a block. 
Similarly to before, if authorities act maliciously (e.g., by proposing a block when they are not allowed) they can be voted out. Specifically, a vote against an authority can be casted at each step and if a majority is reached the authority is removed from the list of legitimate authorities.

As more authorities can propose a block during each step, forks can occur. However, fork likelihood is limited by the fact that each non-leader authority proposing a block delays its block by a random time, hence the leader block is likely to be the first received by all the authorities. If forks happen, the GHOST protocol~\cite{wood2014ethereum} is used, which is based on a block scoring approach: leaders' blocks have higher scores, thus ensuring that forks will be eventually solved.

Figure~\ref{fig:cliq_rotation} shows two consecutive steps and how current leader and authorities allowed to propose blocks change. There are $N = 8$ authorities, hence $N - (N/2 + 1) = 3$ authorities allowed to propose a block at each step, with one of them acting as leader (the bold node in Figure~\ref{fig:cliq_rotation}). In Figure~\ref{fig:cliq_1}, the $a_1$ is the leader while $a_2$ and $a_3$ are allowed to propose blocks. In Figure~\ref{fig:cliq_2},  $a_1$ is not allowed anymore to propose a block (it was in the previous step, so it has to wait $N/2 + 1$ steps), while $a_4$ is now authorised to propose and $a_2$ is the current leader.

\begin{figure}[!tb]
	\centering
	\includegraphics[scale=2.75]{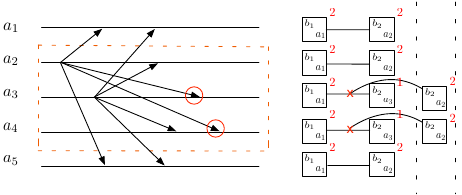}
	\vspace*{-.2cm}
	\itcaption{A fork occurring in Clique. Authority $a_4$ has the block proposed by $a_3$ as second block, while $a_5$ has the block proposed by $a_2$. Eventually, $a_4$ replaces the block proposed by $a_3$ with that proposed by $a_2$ because the latter has a higher score.}
	\label{fig:cliq-fork}
\end{figure}

Regarding the message exchange (see Figure~\ref{fig:cliq}), at each step the leader broadcasts a block and all the authorities directly commit it to the chain. By way of example, Figure~\ref{fig:cliq-fork} depicts a step where the leader authority $a_2$ proposes a new block, as well as the other allowed non-leader authority $a_3$. The former block precedes the latter in the views of $a_1$ and $a_5$, while the opposite occurs for $a_4$ and $a_3$. The resulting fork (see right-hand side of Figure~\ref{fig:cliq-fork}) is easily detected by each authority when the next block is received, as it references a previous block not at disposal of the authority. By relying on the scoring mechanisms (i.e. blocks proposed by leaders win), the GHOST protocol resolves the forks.

% !TEX root = ../../Tesi-1side.tex

\section{Comparison of Aura, Clique and PBFT} \label{s:poaVSpbft}

Previous PoA algorithms are here compared with PBFT, first in terms of consistency and availability properties via the CAP Theorem (Section~\ref{s:safety-liveness}), then of performance (Section~\ref{s:performance}).

\subsection{Consistency and Availability Analysis based on CAP Theorem} 
\label{s:safety-liveness}
Referencing the CAP Theorem introduced in Section \ref{s:consensus-perm-block}, we refine the definitions of CAP properties in the context of permissioned blockchains in order to delving into the CAP-based analysis of the considered algorithms. Moreover we consider a system deployed over the Internet, hence subjected to unforeseeable network delays of variable durations.
In the following, we then assume an eventually synchronous network model where messages can be delayed among correct nodes, but eventually the network starts behaving synchronously and messages will be delivered (within a fixed but unknown time bound, see Section \ref{s:consensus-network}). This model is considered appropriate when designing real resilient distributed systems~\cite{cachin2017blockchains}.

\smallskip
\noindent
\textit{Consistency.}\  
A blockchain achieves consistency when forks are avoided. This property, as reported in Section \ref{s:consensus-perm-block}, is referred to as consensus finality which, in the standard distributed system jargon, corresponds to achieving the \textit{total order} and \textit{agreement} properties of atomic broadcast. The latter is the communication primitive considered as the relevant type of consensus  for blockchains~\cite{cachin2017blockchains}. When consistency cannot be obtained, we have to distinguish whether forks are resolved sooner or later (\textit{eventual consistency}) or they are not (\textit{no consistency}).

\smallskip
\noindent 
\textit{Availability.}\
A blockchain is available if transactions submitted by clients are served and eventually committed, i.e. permanently added to the chain.

\smallskip
\noindent 
\textit{Partition Tolerance.} \ 
When a network partition occurs, authorities are divided into disjoint groups in such a way that nodes in different groups cannot communicate each other.

\begin{figure}[!thbp]
	\centering
	\includegraphics[scale=0.75]{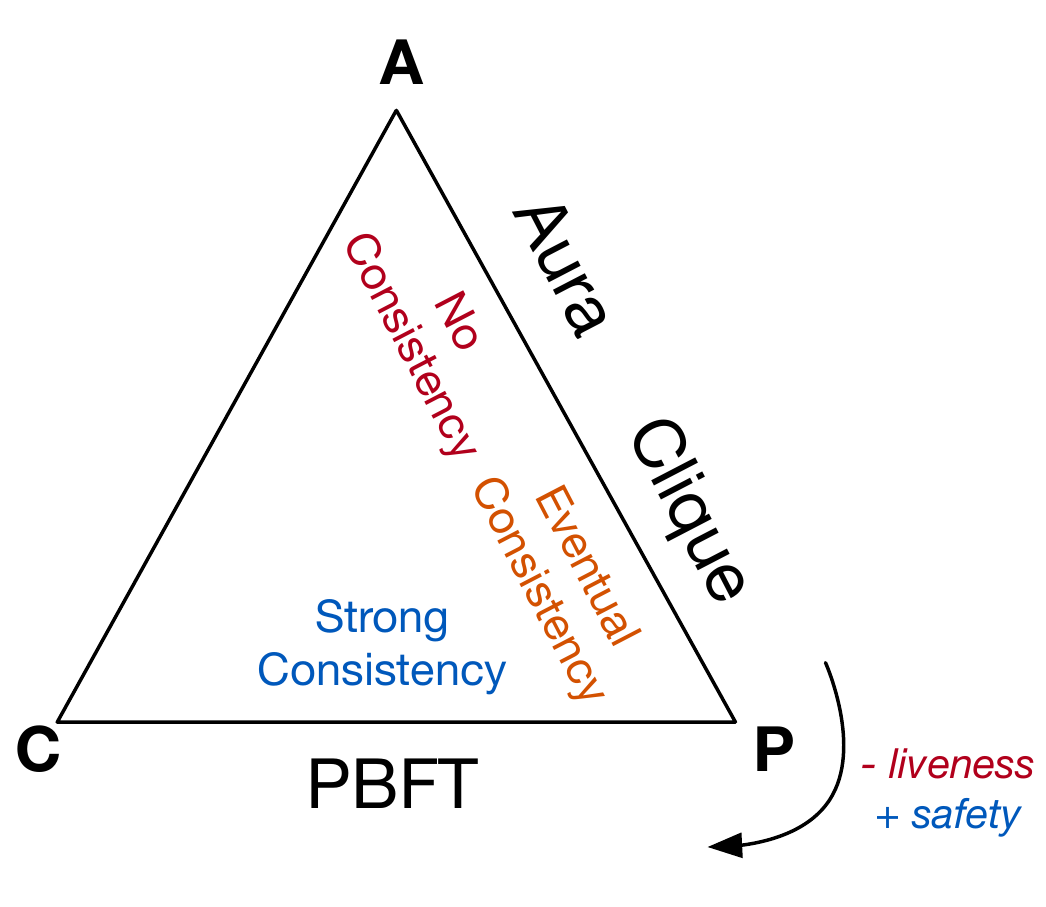}
	\vspace*{-.4cm}
	\itcaption{Classification of Aura, Clique and PBFT according to the CAP Theorem}
	\label{fig:cap-byznt}
\end{figure}

\medskip
\noindent
Therefore, an Internet-deployed permissioned blockchain has to tolerate these adverse situations: (i) periods where the network behaves asynchronously; (ii) a (bounded)  number of Byzantine authorities aiming at hampering availability and consistency. The maximum number of tolerated Byzantine nodes depends on the consensus algorithm: N/2 for PoA, N/3 for for PBFT. Since a blockchain must tolerate partitions, hence CA option is not considered, we analyse the algorithms with respect to CP and AP options. The results of this analysis are reported in Figure~\ref{fig:cap-byznt} and commented in the following. 

\subsubsection{Aura Analysis}
Being based on UNIX synch time, authorities' clocks can drift and become out-of-synch. When authorities are distributed geographically over a wide area, resynchronization procedures cannot be effective due to network eventual synchrony. Hence, there can be periods where authorities do not agree on what is the current step and consequently on the current authority in force. Clocks' skews can be reasonably assumed strictly lower than the step duration, which is in the order of seconds, thus we can have short time windows where two distinct authorities are both considered as leaders by two disjoint sets of authorities, say $\mathcal{A}_1$ and $\mathcal{A}_2$. This can critically affect the consistency of the whole system.

\begin{figure}[!htb]
	\centering
	\includegraphics[scale=1]{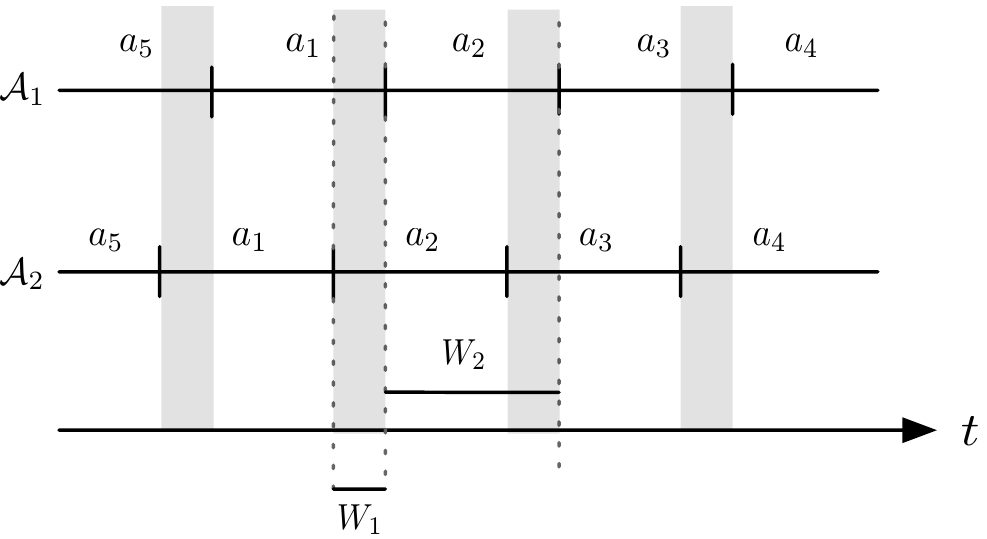}
	\vspace*{-.2cm}
	\itcaption{Example of out-of-synch authorities in Aura (each step for a set of authorities is labelled by the expected leader)}
	\label{fig:aura-desync}
\end{figure}

Let $N_1 = \lvert \mathcal{A}_1 \rvert$ and $N_2 = \lvert \mathcal{A}_2 \rvert$ (where $N_1 + N_2 = N$ and $N$ is an odd number) be the number of authorities in the two sets, respectively. We have a majority of authorities, say $\mathcal{A}_1$, agreeing on who is the current leader. This leads to a situation as depicted in Figure~\ref{fig:aura-desync}: authorities in $\mathcal{A}_1 = \{a_1,a_3,a_5\}$ see steps slightly out of phase with respect to the authorities in $\mathcal{A}_2 = \{a_2,a_4\}$. Indeed, the time windows coloured in grey are those where $\mathcal{A}_1$ disagrees with $\mathcal{A}_2$ on who is the current leader. 
During time window $W_1$, $a_2$ considers itself the leader and sends a block to the other authorities.
$a_2$ is believed to be the leader by the authorities in $\mathcal{A}_2$ but not by those in $\mathcal{A}_1$, hence the former authorities accept its block while the latter ones reject it.
During the time window $W_2$, authorities in $\mathcal{A}_1$ expect $a_2$ to send a block but this does not occur because it has already sent its block for the current step: at the end of $W_2$ authorities in $\mathcal{A}_1$ will vote $a_2$ as malicious and being a majority they will force $a_2$ to be removed.
Therefore, all the remaining authorities in $\mathcal{A}_2$ will be eventually voted out one by one analogously. 
As $\mathcal{A}_2$ is a minority, its authorities are voted out in a number of steps lower than that required to commit enqueued blocks (see Section~\ref{s:poa-aura}), hence there will be no different views of the chain and the consistency is preserved. It is to note that in this case all the authorities are honest.

The same situation can affect consistency when authorities are Byzantine. Let us consider a scenario where there are $B$ malicious authorities, all in $\mathcal{A}_1$ and, differently from before, they do not vote against authorities in $\mathcal{A}_2$. If $B \ge N_1 - N/2$, then a majority is not reached to vote out authorities in $\mathcal{A}_2$, hence the blocks they have proposed achieve finality and are committed to their local chains. This causes a fork that is never resolved: if authorities are not voted out, their blocks are consider as valid and part of the chain.
Therefore, a minority of Byzantine authorities is sufficient to realise this attack and causes \textit{no consistency} in the system.

%%%%%%%%%% PROOF
Indeed, let us consider a set $\mathcal{S}$ with an odd number of elements $N = 2K + 1$, and a partition of such set in two non-empty subsets $\mathcal{S}_1$ and $\mathcal{S}_2$ with cardinality $N_1$ and $N_2$, respectively, such that $\mathcal{S}_1$ is a majority and $\mathcal{S}_2$ a minority, i.e.,

\begin{equation} \label{eq:n1}
K + 1 \le N_1 \le 2K
\end{equation}

\begin{equation*}
1 \le N_2 \le K 
\end{equation*}

\begin{equation*}
N_1 + N_2 = N
\end{equation*}

\noindent We want to prove that it suffices to remove a minority~\footnote{A minority with respect to the set $\mathcal{S}$.} of $B$ elements from $\mathcal{S}_1$ to make it become a minority. Hence, we want to prove that

\begin{equation} \label{eq:B}
\exists B \mid N_1 - B \le K \land B \le K
\end{equation}

\noindent \textit{Proof.} Equation~\ref{eq:B} can be proved by demonstrating that $N_1 - K \le K$. This expression can be written as $N_1 \le 2K$, which is always verified because of Equation~\ref{eq:n1}.
%%%%%%%%%%%%%% FINE PROOF

Anyway, transactions keep being committed over time regardless of what a minority of Byzantine authorities do. Hence, Aura can be classified as an AP system, with no consistency guarantees.

\subsubsection{Clique Analysis}
By design, Clique allows more than one authority to propose blocks with random delays. This permits coping with leaders that could not have sent any block due to either network asynchrony or benign/Byzantine faults.
Resulting forks are anyway resolved by the Ethereum GHOST protocol, hence we have \textit{eventual consistency}.
On the other hand, as the mining frequency of authorities is bounded by $\frac{1}{N/2 + 1}$, a majority of Byzantine authorities is required to take over the blockchain. 
This PoA algorithm can thus be classified as AP, with eventual consistency guarantees.

\subsubsection{PBFT Analysis}
As long as less than one third of nodes are Byzantine, PBFT has been proved to guarantee consistency, i.e. no fork can occur~\cite{castro1999practical}. 
Because of the eventual synchrony of the network, the algorithm can stall and blocks cannot reach finality.  In this case, consistency is preserved while availability is given up. PBFT can then be easily classified as a CP system according to the CAP theorem.

\subsection{Performance Analysis} \label{s:performance}

The analysis here reported is qualitative and only based on how the consensus algorithms work in terms of message exchanging. 
The performance metrics usually considered for consensus algorithms are transaction latency and throughput. 
In the specific case of permissioned blockchains, we measure the latency of a transaction $t$ as the time between the submission of $t$ by a client and the commit of the block including $t$. 
Contrary to CPU intensive consensus algorithms such as PoW, here we can safely assume that latency is communication-bound rather than CPU-bound, as there is no relevant computation involved.
Hence, we can compare the algorithms in terms of the number of message rounds required before a block is committed.
Evaluating the throughput at a qualitative level is much more challenging, as it closely depends on the specific parallelisation strategy (e.g., pipelining) employed by each algorithm implementation.
Thus, we deem more correct to compare throughput performance by the means of proper experimental evaluations that we plan to carry out as future work.

We assess how many message rounds are required for each algorithm in the normal case, i.e. when no condition occurs that makes any corner case to be executed. 
For example, for Aura we do not consider the situation when some authorities suspect the presence of subverted nodes and trigger a voting.

In Aura, each block proposal requires two message rounds: in the first round the leader sends the proposed block to all the other authorities, in the second round each authority sends the received block to all the other authorities.
A block is committed after a majority of authorities have proposed their blocks, hence the latency in terms of message rounds in Aura is $2(N/2 + 1)$, where $N$ is the number of authorities. 

In Clique, a block proposal consists of a single round, where the leader sends the new block to all the other authorities.
The block is committed straight away, hence the latency in terms of message rounds in Clique is $1$.
Such a huge difference between Aura and Clique is due to their different strategies to cope with malicious authorities aiming at creating forks:  Aura waits that enough other blocks have been proposed before committing, Clique commits immediately and copes with possible forks after they occur.
Clique seems to outperform PBFT too, which takes three message rounds to commit a block.

\chapter{Benchmark for permissioned blockchain}\markchapter[Chapter~\thechapter:~Benchmark for permissioned blockchain]
\label{c04:benchmark}
% !TEX root = ../../Tesi-1side.tex
In the previous chapters we outline the difficulty of comparing the rising BFT-based consensus protocols for blockchain systems. These solutions should guarantee atomic broadcast for transactions order, to the detriment of availability or partition-tolerance. However the evaluation of these algorithms is actually challenging in adversarial environment. Indeed to be secure, a protocol should comes out with clear security model and trust assumptions, under which the protocol  proceeds as expected guaranteeing safety and liveness properties.
Actually most of the modern BFT proposals for blockchain comes out without these information, so is difficult to evaluate their employability.

In the information technology is well known that through the experimental evaluation can be only demonstrated the failure of a system \cite{cachin2017blockchains}. Furthermore, by simulating possible attack scenarios can be measured the resiliency against adversarial environments. Thus, for the evaluation of blockchain systems (for the remainder of the chapter this term will be used interchangeably with 'blockchain') we highlight the need of a general purpose benchmark who measure performances under the assumptions above.
\smallskip

In this chapter we propose the implementation of such a benchmark. We move towards a formalisation of blockchain to define a framework for benchmarking and evaluating these algorithms in a more formal approach. Concerning this, we define the evaluation metrics under which measure performances and security. Specifically, to evaluate security, we identify from the general Byzantine model (Section \ref{s:consensus-fail-model}) three most significant attacks for a blockchain system to be exploited in the experimental evaluation.

To analyse the fault tolerance behaviour and hence the resiliency of the consensus algorithms, we propose an experimental approach by taking a real blockchain client and making it Byzantine. By monitoring the behaviours at runtime of the systems, we can then evaluate the impact of Byzantine nodes in real blockchain systems. 
\smallskip

\noindent\textit{Chapter Structure.} Section \ref{s:benchmark-sysmdl} introduces the system model for a permissioned blockchain and the threat model for the benchmark. Section \ref{s:benchmark-bench} introduces the benchmark and the evaluation metrics for measurements. Finally Section \ref{s:benchmark-client} illustrate the practical implementation of a Byzantine node in a real blockchain client.

%\input{chapters/03_01_pascal-related}
% !TEX root = ../../Tesi-1side.tex
\section{System and Threat Model}\label{s:benchmark-sysmdl}
% Questo commento va nel capitolo 2, quando si parla della necessità di avere un modello generico per algoritmi di consenso su blockchain.
% 1. To build a security model for blockchain we should consider all possible elements in the system s.t. the network, the availability of synchronized clocks and the expected (mis-)behavior of nodes e.g. the more generic trust assumption is f < n/k faulty-nodes in the system.
In this section is introduced the system model for permissioned blockchain. We assume a \textit{partially synchronous} communication network (see Section \ref{s:consensus-network}). A peer-to-peer distributed system of authenticated nodes communicate over the network and keep a common state update. This state is essentially a replicated data structure shared in memory between replicas, i.e. \textit{blockchain}. The blockchain is a collection of transactions organised in blocks, it can be referenced as a ledger of blocks where each one is cryptographically linked to the previous by means of its hash. 

\begin{figure}[h]
 \centering
 \includegraphics[scale=0.6]{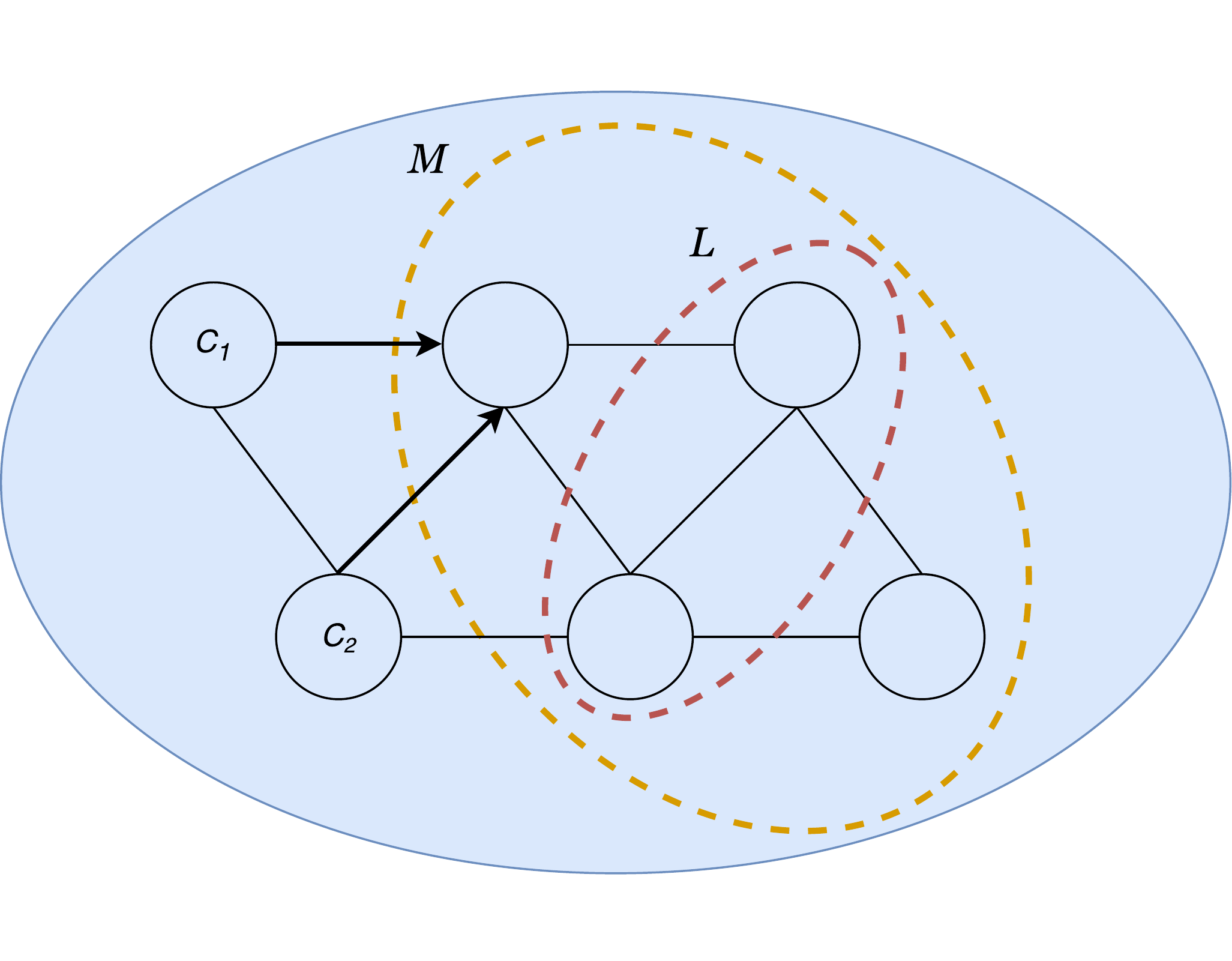}
 \itcaption{Example of blockchain network with six nodes which maintains the ledger. There are four miners of which two possible leaders. There are also two clients which submit transactions (bold arrows). Note that a node of the network can be a miner and a client simultaneously.}
 \label{fig:mesh1}
\end{figure}

A transaction represent an asset exchange between two parties, secured by asymmetrical cryptography, while blocks are essentially collections of transactions with a determined size. In this scenario we identify two class of nodes, \textit{miners} and \textit{clients}. The clients are nodes in charge to submit transactions, while miners are responsible for accepting correct transactions and collect them in blocks. We define the following property for the transaction validation:
\paragraph{External Validity \textit{(safety)}} Verifies the correctness of transactions broadcasted through the network from each client. Only valid transactions can be accepted by the protocol and committed to the blockchain. The property has been defined as \textit{external validity} by Cachin et al.\cite{DBLP:journals/iacr/CachinKPS01} and can be formalized as an external predicate \textit{V} deterministically computable by every node locally. This property can be formalized as follows\cite{cachin2017blockchains}:
  \theoremstyle{definition}
  \newtheorem{definition}{Definition}
  \begin{definition}{\textbf{(Transaction Validation)}}
    \label{def:1}
    \textit{Given a deterministic predicate V s.t. $V$ $\rightarrow \{True, False\}$, and given two correct nodes of the blockchain $n_1, n_2$ s.t in atomic broadcast they have committed the same sequence of transactions $\mu$.
    Than $n_1$ obtain, $\forall tx\in\mu$, V(tx)=True iff $\forall tx\in\mu$, $n_2$ obtain V(tx)=True}
  \end{definition}

\begin{figure}[h]
 \centering
 \includegraphics[scale=0.7]{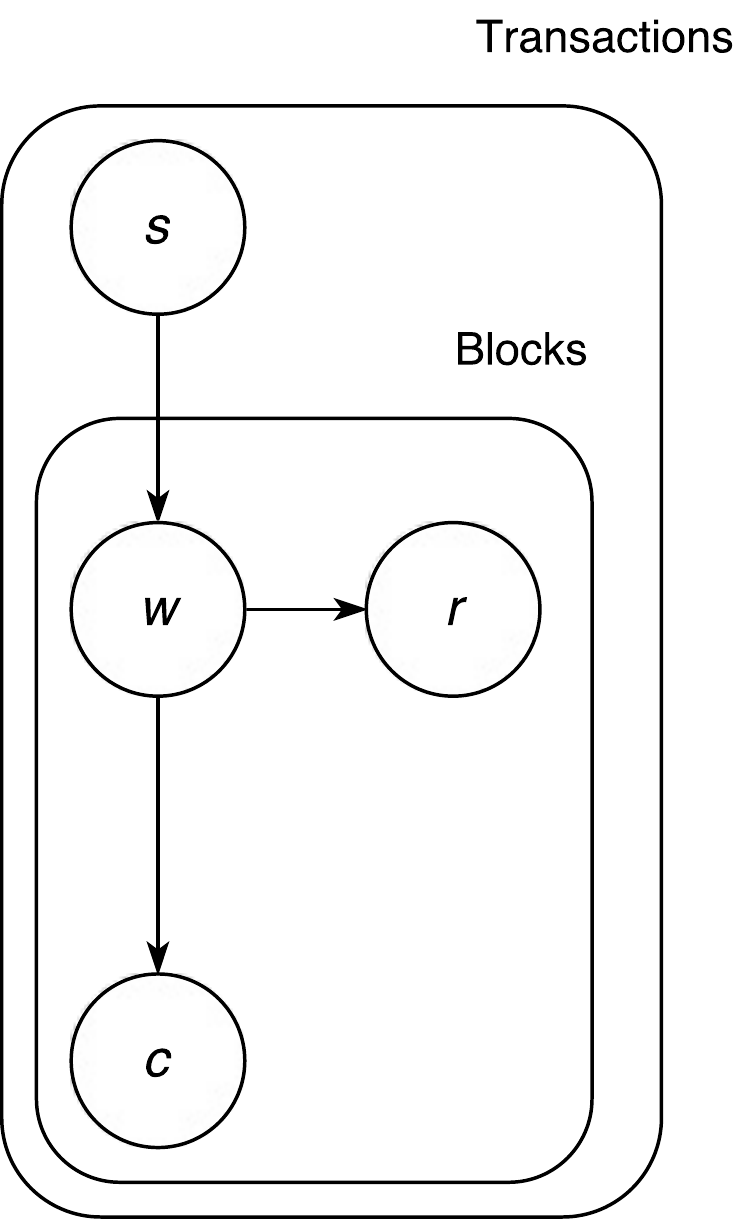}
 \itcaption{Transactions are Submitted by clients to the network. Then, once these transactions are collected in blocks, they change the state with the relative block. Each block waits for consensus in a Wait state or can be Refused. When the miners reach consensus a block will be Committed.}
 \label{fig:mesh2}
\end{figure}

After the validation, each miner keep a list of pending transactions and create the relative blocks to be proposed as next of the blockchain. While a block, and so the included transactions, is not committed immediately on the blockchain, initially it is kept in a waiting state. Thus, to guarantee atomicity on the transactions order, a consensus protocol is run between miners to establish the next block to be committed in the shared state. We identify a special class of nodes, i.e. \textit{leaders}, which are miners authorised to propose a block at a certain point in time. Figure \ref{fig:mesh1} illustrates an example of our modelled blockchain network. 

When miners achieve consensus the current leader propose a new block. This can be accepted and committed to the blockchain by all the nodes of the network, or refused (Figure \ref{fig:mesh2} is illustrated the life-cycle of transactions and blocks).
\smallskip

The blockchain system can be splitted into two layers. Miners of the network executes, in a first layer, an algorithm for the collection of transactions submitted by clients and for the management of blocks. This level is supported by a second layer which runs the consensus protocol. This modules are connected each other, granting the execution of transactions and the update of the shared ledger. 

To formalise the functioning of the first layer we model the scenario described above as follows.

\medskip
\noindent Let $B$ be a blockchain system, $B:=<\mathcal{M}, \mathcal{L}, \mathcal{C}, \Gamma, \Omega, \alpha, \beta>$ where:
\begin{itemize}
    \item $\mathcal{M}$ is the set of miners
    \item $\mathcal{L}$ is the set of possible leaders, $\mathcal{L}\subseteq \mathcal{M}$
    \item $\mathcal{C}$ is the set of clients
    \item $\Gamma$ is the set of transactions
    \item $\Omega$ is the set of blocks
    \item $\alpha$ is the consensus algorithm adopted, it determine:
    \begin{enumerate}
        \item when a new leader is elected from miners:
        \begin{itemize}
            \item if the current leader is byzantine (PBFT);
            \item at each block (Mining Rotation);
        \end{itemize}
        \item the subset of miners which can be leader:
        \begin{itemize}
            \item all miners, $\mathcal{L}=\mathcal{M}$ (PBFT);
            \item a subset of miners, $\mathcal{L}\subseteq \mathcal{M}$ (Mining Rotation).
        \end{itemize}
    \end{enumerate}
    \item $\beta:(\Gamma$ x $\mathcal{M})\rightarrow \Omega$, is a function executed by miners, which takes in input the submitted transactions and return a new block. 
\end{itemize}

Referencing to the semantics defined, in the Algorithm \ref{alg:1}  we propose the standardisation of the first layer, where miners collect transactions and prepare blocks. This algorithm rely on perfect links for communications and on a reliable broadcast module to guarantee the correct delivery of transaction. Moreover the algorithm rely on a consensus protocol module to manage the leaders' proposals.

\algdef{SE}[EVENT]{UEvent}{EndUEvent}[1]{\textbf{upon event}\ #1\ \algorithmicdo}{\algorithmicend\ \textbf{event}}%
\algtext*{EndUEvent}
\algdef{SE}[EVENT]{uEvent}{EnduEvent}[1]{\textbf{upon}\ #1\ \algorithmicdo}{\algorithmicend\ \textbf{event}}%
\algtext*{EnduEvent}
\algdef{SE}[EVENT]{Func}{EndFunc}[2]{\textbf{function}\ #1\ \algorithmicreturn\ #2\ }{\algorithmicend\ \textbf{event}}%
\algtext*{EndFunc}
\begin{algorithm}[!ht]
\caption{First layer blockchain}
\textbf{Uses: }\\
PerfectPointToPointLink $pl$;\\
ReliableBroadcast $rb$;\\
ConsensusProtocol $\alpha$.\\
\begin{algorithmic}[1]
\UEvent{$< Init >$}
    \State$submitted:=0$
    \State$waiting:=0$
    \State$refused:=0$
    \State$committed:=0$
    \State$BLOCK\_SIZE:=k$  \Comment{Predefined block size}
\EndUEvent
\UEvent{$< rb, Deliver|c,tx >$}
    \If{$external\_validity(tx)$}
        \State$submitted:=submitted\cup\{tx\}$
    \Else
        \State\textit{\textbf{trigger}} $< pl, Refused|c,tx >$
    \EndIf
\EndUEvent
\uEvent{$|submitted|==BLOCK\_SIZE$}
    \State$block\_tx=submitted$
    \State$block=\beta(block\_tx)$
    \State$submitted:=submitted\backslash\{block\_tx\}$
    \State$waiting:=waiting\cup\{block\}$
    \State\textit{\textbf{trigger}} $< rb, Broadcast|self,block >$
\EnduEvent
\UEvent{$< rb, Deliver|m,block >$}
    \If{$verify(block)$}    \Comment{Function for block validity verification}
        \State\textit{\textbf{trigger}} $< \alpha, Propose|m,block >$
    \Else
        \State$refused:=refused\cup\{block\}$
        \State$waiting:=waiting\backslash\{block\}$
    \EndIf
\EndUEvent
\UEvent{$< \alpha, Deliver|block,state >$}
    \If{$state==accepted$}
        \State$committed:=committed\cup\{block\}$
        \State$waiting:=waiting\backslash\{block\}$
    \ElsIf{$state==refused$}
        \State$refused:=refused\cup\{block\}$
        \State$waiting:=waiting\backslash\{block\}$
    \EndIf
\EndUEvent
\Func{$\beta(block\_tx)$}{$\Omega$}
    \ForAll{$tx\in block\_tx$}
        \State$block:=block\cup\{tx\}$
    \EndFor
    \State\textbf{return} $block$
\EndFunc
\end{algorithmic}
\label{alg:1}
\end{algorithm}

%%%%%%%%%%%%%%%%%%%%%%%%%%%%%%%%%%%%%%%%%%%%%%%%%%%%%%%%
\subsection{Blockchain Properites}\label{s:benchmark-sysmdl-prop}
The atomic broadcast is the most feasible consensus in permissioned blockchain. However, besides its fundamental properties (see Section \ref{s:consensus-atomic-bcast}), the peculiarities of blockchain require the analyse of additional properties of interest. We list them in the following.

Given a blockchain system with a set of nodes such that: $M:=\{m_1,.., m_i\}$ miners, and $C:=\{c_1,.., c_j\}$ clients, we identify as pending transaction $tx\in \Gamma$ and a pending block $b\in \Omega$ s.t. these properties are valid:
\paragraph{Integrity \textit{(safety)}}
  \begin{itemize}
     \item \textit{No duplication}: $tx$ is not committed to the blockchain more than once.
     \item \textit{No creation}: If $tx$ is committed to the blockchain, than it was previously broadcasted by a correct miner $m_j$.
  \end{itemize}

\paragraph{Validity \textit{(safety)}} If a correct node commits a transaction $tx$ to the blockchain, in a block $b$, then $tx$ is committed, in the same block $b$, by every correct node.

\paragraph{Finality \textit{(safety)}} In BFT-based blockchain, \textit{conensus finality} is the possibility of encounter forks during the chain construction. Roughly speaking finality is reached if all valid blocks appended to the blockchain at some point in time, will be never removed from the blockchain (see Section \ref{s:consensus-perm-block} for a formal definition).
  
\paragraph{Termination \textit{(liveness)}} For every transaction $tx$, if at least one correct node commits $tx$. then all correct nodes (eventually) commit $tx$.

%%%%%%%%%%%%%%%%%%%%%%%%%%%%%%%%%%%%%%%%%%%%%%%%%%%%%%%%

\subsection{Threat Model}\label{s:benchmark-sysmdl-att}

Assessing the security of blockchain consensus protocols is a challenging issue due to the variety of attacks faced by blockchain system. Defining an attacker model is a first step toward a comprehensive experimental evaluation of the resiliency of blockchain consensus protocols. 

We model an attacker by identifying three more significant attacks for a blockchain system. As an approach, we started from the Byzantine failure model~\cite{CRISTIAN1995158} (Section \ref{s:consensus-fail-model}), where failures may occur or nodes may not behave in the proper manner because subverted. Specifically, three classes of failures are there identified: \textit{omission failure}, \textit{timing failure} and \textit{Byzantine failure}. These classes are also specialised for sub-class of authenticated protocols, likewise digitally signing a block or a transaction in a blockchain system.
%We assume an \textit{authentication-detectable Byzantine failures} model which is a subclass of Byzantine failures where is employed the use of a message authentication protocol, i.e. digital signatures of blocks and transactions, to detect the corruption of messages. 

%In this paper we refer to the authentication-detectable Byzantine failures model simply as Byzantine model. This model holds the omission and timing failures where messages can be delayed or lost, and the Byzantine behaviour where a message may be altered.

%We design three possible attack scenarios for a \textit{miner}, a special node authorized to dispatch blocks through the network, then we evaluate algorithms resiliency against these attacks. 
\paragraph{Attacker model} 
We assume a blockchain system formed by \textit{n} independent nodes participating to the consensus algorithm with at most \( f < \frac{n}{k}, (k = 2,3,..) \) Byzantine faulty nodes. We identify three attacks: 
\begin{itemize}
\item \textit{(i) leader misbehaviour}, a timing fault due to a miner broadcasting inconsistent blocks at the same time; 
\item \textit{(ii) forgery}, a Byzantine fault of a miner trying to forge fake transactions and/or blocks; 
\item \textit{(iii) Denial of Service}, an omission fault due to a miner avoiding the signing or broadcasting of a transaction.
\end{itemize}
An attacker aims at compromising \textit{safety} and \textit{liveness} of an algorithm by exploiting these attacks. Specifically, by referring to the evaluation properties reported before, we have that (i) leader misbehaviour violates finality; (ii) forgery violates Integrity, validity and external validity; (iii) Denial of Service violates termination. 
% !TEX root = ../../Tesi-1side.tex

\section{Benchmark}\label{s:benchmark-bench}
In this section we define a general methodology to standardise the evaluation of permissioned blockchain systems. To benchmark such a system we should focus on the underling consensus algorithm. Usually for the evaluation metrics of consensus the most relevant properties to be considered are (i) \textit{performance}and, (ii) \textit{security}.

To evaluate performances we consider throughput and latencies of transactions and, scalability. Furthermore, because blockchains are \textit{asynchronous services} \cite{dinh2017blockbench}, transactions submitted to the system are not processed immediately. To measure the latency of transactions we refer to the time between the submission and the commit of the block including the transaction.

To evaluate the impact of the attacks listed  in the system model, Section \ref{s:benchmark-sysmdl-att}, we refer to the following properties.

\subsection*{Evaluation Metrics}
\ite
\item \textit{Throughput}: measured as the number of transactions successfully committed per second.
\item \textit{Latency}: measured as the time between the transaction submission and the commit of the relative block. 
\item \textit{Scalability}: measured as the changes of throughput and latency, increasing the number of nodes of the network.
\item \textit{Safety}: identified with the Integrity, Validity and Finality properties listed in Section \ref{s:benchmark-sysmdl-prop}  
\item \textit{Liveness}: identified with the Termination property listed in Section \ref{s:benchmark-sysmdl-prop}
\eti

For each attack we propose to measure a sort of \textit{index-of-tolerance} which identify how the threat impacts the metrics, i.e. a performance breakdown or violation of safety and liveness.
% !TEX root = ../../Tesi-1side.tex

\section{Byzantine client for blockchain}\label{s:benchmark-client}
For evaluation purposes, whereas possible, we illustrate the best practice to integrate a Byzantine client in a real blockchain framework. Most of these blockchain systems follow a similar structure which covers the same (or similar) actors. 

The Byzantine client we propose to integrate should simulate the attacker model given in Section \ref{s:benchmark-sysmdl-att}. Each attack impacts some properties and functions of the architectural implementation.

We list below the best practice to follow for achieving this task and identify the correct functions to modify in a blockchain framework.

\enu
\item Use the system logs to understand the the behaviour of the relevant actors. Specifically we suggest to use the \text{trace} level of the logger and focus on (i) transactions, (ii) miners and, (iii) consensus engine;
\item Allow a blockchain client to run as Byzantine from the CLI. From the source code add the flag option to the CLI manager;
\item Once identified the miner options, update their configurations to be prepared for a Byzantine set-up. Indeed this functions usually manage the miner behaviours, i.e. the block creation, the block sealing and dispatching. Likewise a Byzantine miner must be provided of specific functions to simulate the attacks above;
\item Update the consensus engine module which interact with the functions at 3. This module needs to be prepared at Byzantine behaviours and, because it manage the achievement of consensus, new functions for the possible attacks management are required.
\une

Currently we have implemented a demo of this Byzantine client inside Parity Ethereum's client. We simulate the denial-of-service attack forcing a miner to ignore the block sealing procedure. For the implementation details we refer to the Appendix \ref{a:byz-client}.

%%%%%%%%%%%%%%%%%%%%%%
% PART 3: CONCLUSION %
%%%%%%%%%%%%%%%%%%%%%%

\chapter{Conclusions and Future Directions}\markchapter[Chapter~\thechapter:~Conclusions and Future Directions]
\label{c05:conclusion}
% !TEX root = ../../Tesi-1side.tex

In this thesis we approached some techniques to analyse the permissioned blockchain consensus protocols. Specifically we studied the security requirements relative to safety and liveness properties and we analyse their performance guarantees. The thesis brings two main contributions. The first contribution is a qualitative analysis of two consensus algorithms against CAP (consistency, availability and partition tolerance) properties. The second contribution is a formalisation of permissioned blockchains, for defining a benchmarking framework and evaluating the consensus algorithms with a more formal approach.

\bigskip
In the first contribution we derive the functioning of two prominent consensus algorithms for permissioned blockchains based on the PoA paradigm, namely Aura and Clique. 
We provide a qualitative comparison of them with respect to PBFT in terms of consistency, availability and performance, by considering a deployment over the Internet where the network is realistically modelled  as eventually synchronous rather than synchronous. By applying the CAP Theorem, we claim that in this setting PoA algorithms can give up consistency for availability when considering the presence of Byzantine nodes.
This can prove to be unacceptable in scenarios where the integrity of transactions has to be absolutely kept (which is likely to be the actual reason why a blockchain-based solution is used). On the other hand, PBFT keeps the blockchain consistent at the cost of availability, even when the network behaves temporarily asynchronously and Byzantine nodes are present; this behaviour is much more desirable when data integrity is a priority. Despite one of the most praised advantages of PoA algorithms is their performance, our qualitative analysis shows that in terms of latency the expected loss of PBFT is bounded, and can be offset by the gain in consistency guarantees.

\smallskip
In the second contribution we detailed a formal model for permissioned blockchain. Specifically we outline the most general model for such a scenario, to be as feasible as possible in real context. We assume a partially synchronous network which is subject to the most general Byzantine fault model. Thus we profile a possible attacker model that can be exploited over a blockchain network. To measure security, we assess the evaluation metrics in correct behaviour and under the attacks listed in the system model, Section \ref{s:benchmark-sysmdl-att}. Specifically we propose to evaluate the impact of the attacks to the specified evaluation metrics, and for this purpose we integrate a first prototype of a Byzantine client node in a real blockchain, i.e. Parity.

\bigskip
As future work, we plan to deepen the analysis of PoA algorithms by engaging further reverse engineering tasks and thorough experimental evaluations. The final goal is to validate and possible revise our claims on the availability and consistency guarantees of PoA and PBFT, by implementing the adverse scenarios we envisioned in Section~\ref{s:safety-liveness}. Furthermore with the specified benchmark guidelines, we want to create a real benchmarking architecture to collect performance measurements, both transaction latency and throughput, and to test scalability with respect to varying input transaction rates and number of nodes/authorities. For this purpose we aim to experiment the consensus technologies by testing the impact of the Byzantine client in consensus algorithms and outline an evaluation metric for security, i.e.  \textit{index-of-tolerance}, to measure how a specific algorithm is resilient to an adversarial context. 

%%%%%%%%%%%%%%%%%%%%%%%%%%%%%%%%%%%%%%%%%%%%%%%%%%%%%%%%%%%%%%%%%%%%%%%%%%%%%%
%% Per far comparire il numero di capitolo in testa alla pagina sinistra:
%%   \chapter{Introduction}
%%   \markchapter
%%
%% Per far comparire nome e numero del capitolo in testa alla pagina sinistra:
%%   \chapter{Introduction}
%%   \markchapter[\thechapter~Introduction]
%%
%% Per far comparire nome e numero di sezione in testa alla pagina destra:
%%   \marksection{Some First Level Section}
%%   \section{Some First Level Section}
%%   \marksection{Some First Level Section}
%%%%%%%%%%%%%%%%%%%%%%%%%%%%%%%%%%%%%%%%%%%%%%%%%%%%%%%%%%%%%%%%%%%%%%%%%%%%%%

%%%%%%%%%%%%%
% Citazione %
%%%%%%%%%%%%%
\cleardoublepage
\section*{}
\vspace*{6.5cm}
\begin{flushright}
	\emph{''Life is what happens to you while you are making other plans.''}
	
	\smallskip
	John Lennon
\end{flushright}

%%%%%%%%%%%%%%%
%% Appendici %%
%%%%%%%%%%%%%%%
\begin{appendices}
	\addtocontents{toc}{\protect\setcounter{tocdepth}{1}}
	\makeatletter
	\addtocontents{toc}{%
		\begingroup
		\let\protect\l@chapter\protect\l@section
		\let\protect\l@section\protect\l@subsection
	}
	\makeatother
	\chapter{Parity Byzantine client} \label{a:byz-client}
	% !TEX root = ../../Tesi-1side.tex
We propose the integration of a Byzantine client in Parity blockchain. It is the Ethereum's client based on the programming language Rust. Beyond the possibility of create a proper client on the public Ethereum blockchain, Parity offers the possibility to realise private permissioned blockchain based on lighter consensus engines. We integrate our client following the steps in Section \ref{s:benchmark-client}. Here we show the implementation details for each of the steps.

\bigskip
First of all, once identified the flow of transactions and miners we integrate the possibility of run a node as Byzantine by updating the configurations of the CLI. More precisely we add the boolean flag \colorbox{silver}{\texttt{{\color{red} byzantine}}} by updating the file at \texttt{$\sim$ parity/cli/mod.rs} as follows.

\vspace{0.5cm}

\lstset{
  language={myLang},
  basicstyle=\small\ttfamily, % Global Code Style
  captionpos=b, % Position of the Caption (t for top, b for bottom)
  extendedchars=true, % Allows 256 instead of 128 ASCII characters
  tabsize=2, % number of spaces indented when discovering a tab 
  columns=fixed, % make all characters equal width
  keepspaces=true, % does not ignore spaces to fit width, convert tabs to spaces
  showstringspaces=false, % lets spaces in strings appear as real spaces
  breaklines=true, % wrap lines if they don't fit
  frame=trbl, % draw a frame at the top, right, left and bottom of the listing
  frameround=tttt, % make the frame round at all four corners
  framesep=4pt, % quarter circle size of the round corners
  %numbers=left, % show line numbers at the left
  %numberstyle=\tiny\ttfamily, % style of the line numbers
  backgroundcolor=\color{backcolour},
  commentstyle=\color{eclipseGreen}, % style of comments
  keywordstyle=[1]\color{eclipsePurple}, % style of keywords
  keywordstyle=[2]\color{amber(sae/ece)},
  keywordstyle=[3]\color{eclipseBlue},
  keywordstyle=[4]\color{alizarin},
  stringstyle=\color{dollarbill}, % style of strings
}

\begin{lstlisting}[caption=Byzantine Flag integration.]
...
// -- Sealing/Mining Options
flag_stratum: bool = false,
	or |c: &Config| Some(c.stratum.is_some()),
flag_stratum_interface: String = "local",
	or |c: &Config| otry!(c.stratum).interface.clone(),
flag_stratum_port: u16 = 8008u16,
	or |c: &Config| otry!(c.stratum).port.clone(),
flag_stratum_secret: Option<String> = None,
	or |c: &Config| otry!(c.stratum).secret.clone().map(Some),
flag_byzantine: bool = false,
	or |c: &Config| otry!(c.mining).byzantine.clone()
...
\end{lstlisting}

\vspace{0.5cm}

\begin{lstlisting}[caption=Byzantine Flag for Mining options.]
...
struct Mining {
	author: Option<String>,
	engine_signer: Option<String>,
	force_sealing: Option<bool>,
	reseal_on_uncle: Option<bool>,
	reseal_on_txs: Option<String>,
	reseal_min_period: Option<u64>,
	...
	...
	tx_queue_strategy: Option<String>,
	tx_queue_ban_count: Option<u16>,
	tx_queue_ban_time: Option<u16>,
	remove_solved: Option<bool>,
	notify_work: Option<Vec<String>>,
	refuse_service_transactions: Option<bool>,
	byzantine: Option<bool>,
}
...
\end{lstlisting}

\vspace{0.5cm}

Then to update the miners configurations we load the byzantine flag in the \texttt{miner\_options()} functions at line \texttt{517} of the file \texttt{$\sim$parity/configuration.rs} adding the following option:

\vspace{0.5cm}

\begin{lstlisting}
byzantine: self.args.flag_byzantine,
\end{lstlisting}
\vspace{0.5cm}
hence we add the required miner configurations by editing the file at \\\texttt{$\sim$ethcore/src/miner/miner.rs} as follows.

\newpage

\begin{lstlisting}[caption=Miner behaviour configurations.]
...
pub struct MinerOptions {
	pub new_work_notify: Vec<String>,
	/// Byzantine node.
	pub byzantine: bool,
	/// Force the miner to reseal, even when nobody has asked for work
	pub force_sealing: bool,
	...
	...	
	pub tx_queue_gas_limit: GasLimit,
	pub tx_queue_banning: Banning,
	pub refuse_service_transactions: bool,
}
...
\end{lstlisting}

\vspace{0.5cm}
\begin{lstlisting}[caption=Miner default behaviour configurations.]
...
impl Default for MinerOptions {
	fn default() -> Self {
		MinerOptions {
			new_work_notify: vec![],
			byzantine: false,
			force_sealing: false,
			...
			...
			tx_queue_banning: Banning::Disabled,
			refuse_service_transactions: false,
		}
	}
}
...
\end{lstlisting}

\newpage

\begin{lstlisting}[caption=Byzantine function notifier and If statement for Byzantine client.]
...
	fn is_byzantine(&self) -> bool {
		self.options.byzantine || !self.notifiers.read().is_empty()
	}
...
...
/// Attempts to perform internal sealing (one that does not require work) and handles the result depending on the type of Seal.
	fn seal_and_import_block_internally(&self, chain: &MiningBlockChainClient, block: ClosedBlock) -> bool {
		if !block.transactions().is_empty() || self.forced_sealing() || Instant::now() > *self.next_mandatory_reseal.read() {
			trace!(target: "miner", "seal_block_internally: attempting internal seal.");

			// ADD IF CONDITION FOR match with byzantine (controll with function self.is_byzantine())
			let seal_function = if self.is_byzantine() {
				self.engine.generate_seal_byzantine(block.block())
			} else {
				self.engine.generate_seal(block.block())
			};

			...
			...
\end{lstlisting}
\vspace{1cm}
Once implemented the miner options we update the consensus engine, to simulate the DoS attack. Because Parity provide the possibility to integrate different consensus engines we firstly add the generic function to the interface (\texttt{$\sim$ethcore/src/engines/mod.rs}) and then we implement this function in our reference consensus, i.e. Aura Proof-of-Authority.

\newpage
\begin{lstlisting}[caption=Byzantine sealing function. Exploitation of DoS attack.]
...
/// Byzantine seal. No blocks will be sealed.
/// Edited by @StefanoDeAngelis.
fn generate_seal_byzantine(&self, block: &ExecutedBlock) -> Seal {
	if !self.can_propose.load(AtomicOrdering::SeqCst) { return Seal::None; }

	let header = block.header();
	let step = self.step.load();
	
	let active_set;
	...
	...
	if is_step_proposer(validators, header.parent_hash(), step, header.author()) {
		trace!(target: "engine", "generate_seal: Byzantine proposer for step {}. No block sealed.", step);
	} else {
		trace!(target: "engine", "generate_seal: {} not a proposer for step {}.",
			header.author(), step);
	}
	Seal::None
}
...
\end{lstlisting}

	\chapter{Consensus Finality} \label{a:finality}
	% !TEX root = ../../Tesi-1side.tex
\emph{Consensus Finality} guarantees strong consistency on blockchain among replicas. Essentially it describes the possibility of having forks. In distributed systems is proved that for guaranteeing strong consistency two properties are required: \textit{agreement} and \textit{total order}. For demonstrating that finality strictly implies strong consistency, we reproduce two scenarios of forks which differently violates agreement and total order.

\begin{figure*}[!hbpt]
	\centering
	\subfigure[No Agreement]{
		\label{fig:proof_1}
		\includegraphics[width=7.5cm]{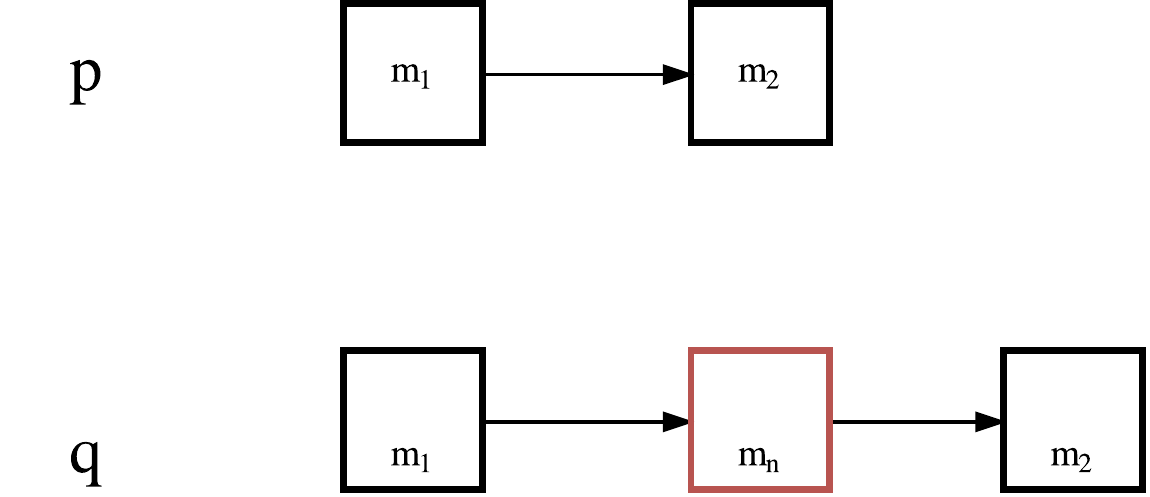}
	}\qquad
	\subfigure[No Total Order]{
		\label{fig:proof_2}
		\includegraphics[width=7.5cm]{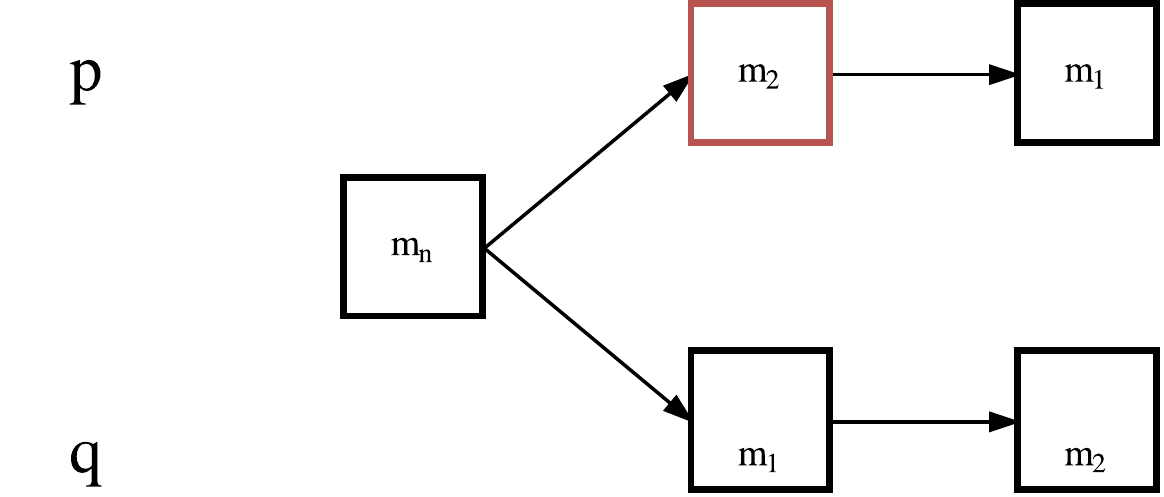}
	}
	\itcaption{Fork scenarios that violate agreement and total order.}
	\label{fig:fin-proof}
\end{figure*}

We consider two correct nodes $p$ and $q$, which hold a blockchain. In \figref{fig:proof_1} agreement is not guaranteed because $q$ have different blocks to the blockchain with respect to $p$. Indeed \figref{fig:proof_2} shows a case in which total order is violated, nodes have the same blocks but with different order. In both scenarios there is a fork between the blockchains of the two participants that violate consequently the finality property. Therefore we can define the following theorem:

\begin{theorem}
A blockchain system achieve finality if and only if agreement and total order properties are guaranteed.
\end{theorem}

\begin{corollary}
If a blockchain system do not achieve finality, it can only guarantee eventual consistency.
\end{corollary}
	
	\addtocontents{toc}{\endgroup}
\end{appendices}

%\markappendix[\thechapter~List of Tables]
%\clearpage\addcontentsline{toc}{chapter}{List of Tables}\listoftables

\markappendix[\thechapter~List of Figures]
\clearpage\addcontentsline{toc}{chapter}{List of Figures}\listoffigures

%%%%%%%%%%%%%%%%%%%%%%%%%%%%%%%%%%%%%%%%%%%%%%%%%%%%%%%%%%%%%%%%%%%%%%%%%%%%%%
%% Per far comparire il numero di appendice in testa alla pagina sinistra:
%%   \chapter{Some Appendix}
%%   \markappendix
%%
%% Per far comparire nome e numero di appendice in testa alla pagina sinistra:
%%   \chapter{Some Appendix}
%%   \markappendix[\thechapter~Some Appendix]
%%%%%%%%%%%%%%%%%%%%%%%%%%%%%%%%%%%%%%%%%%%%%%%%%%%%%%%%%%%%%%%%%%%%%%%%%%%%%%

%%%%%%%%%%%%%%%%%%
%% Bibliografia %%
%%%%%%%%%%%%%%%%%%
\backmatter
\raggedright
\begin{small}
	\addcontentsline{toc}{chapter}{Bibliography}
	\bibliographystyle{plain}
	\bibliography{References}
\end{small}

%%%%%%%%%%%%%%%%%%%%%%%%%%%%%%%%%%%%%%%%%%%%%%%%%%%%%%%%%%%%%%%%%%%%%%%%%%%%%%
%% Per far comparire la bibliografia nell'indice:
%% 1) generare con bibtex la versione definitiva del file thesis.bbl
%% 2) ridenominare thesis.bbl in thesis-bbl.tex (per evitare che un run
%%    accidentale di bibtex lo sovrascriva)
%% 3) aggiungere nel file thesis-bbl.tex dopo \begin{thebibliography}{...} la
%%    riga:
%%          \addcontentsline{toc}{chapter}{Bibliography}
%% 4) sostituire in questo file i due comandi
%%          \bibliographystyle{...} e \bibliography{...}
%%    con
%%          \input{thesis-bbl}
%%%%%%%%%%%%%%%%%%%%%%%%%%%%%%%%%%%%%%%%%%%%%%%%%%%%%%%%%%%%%%%%%%%%%%%%%%%%%%

\end{document}